\documentclass[useAMS,usenatbib]{mnras}

\usepackage[usenames,dvipsnames,svgnames,table]{xcolor}
\usepackage{graphicx}
\usepackage[compatibility=false]{caption}
\usepackage{subcaption}
\usepackage[T1]{fontenc}
\usepackage{times}
\usepackage{aecompl}
\usepackage{comment} 

\usepackage{amssymb} 
\usepackage{amsmath} 
\usepackage{bbding} 
\usepackage{wasysym} 

\usepackage{threeparttable} 

\title{Radio Monitoring of Protoplanetary Discs}
\author[Ubach et al.]
{C.~Ubach$^{1,2}$\thanks{E-mail: cubach@nrao.edu},
S.~T.~Maddison$^{2}$, C.~M.~Wright$^{3}$, D.~J.~Wilner$^{4}$, D.~J.~P.~Lommen$^{5}$,
\newauthor  B.~Koribalski$^{6}$ \\
$^{1}$National Radio Astronomy Observatory, 520 Edgemont Road, Charlottesville VA 22903-4608, USA \\
$^{2}$Centre for Astrophysics \& Supercomputing, Swinburne University of Technology, H30, P.O. Box 218, Hawthorn, VIC 3122, Australia\\
$^{3}$School of Physical, Environmental and Mathematical Sciences, UNSW@ADFA, Canberra ACT 2600, Australia\\
$^{4}$Harvard-Smithsonian Center for Astrophysics, 60 Garden Street, 02138 Cambridge, MA, USA\\
$^{5}$Hwa Chong Institution, 661 Bukit Timah Road, Singapore 269734 \\
$^{6}$CSIRO Astronomy \& Space Science, Australia Telescope National Facility (ATNF), PO Box 76, Epping NSW 1710, Australia\\
}

\begin{document}

\pagerange{\pageref{firstpage}--\pageref{lastpage}} \pubyear{2015}

\maketitle

\label{firstpage}

\begin{abstract}

Protoplanetary disc systems observed at radio wavelengths often show excess
emission above that expected from a simple extrapolation of thermal dust
emission observed at short millimetre wavelengths. Monitoring the emission
at radio wavelengths can be used to help disentangle the physical
mechanisms responsible for this excess, including 
free-free emission from a wind or jet, and chromospheric emission
associated with stellar activity.
%
We present new results from a radio monitoring survey conducted with Australia Telescope Compact Array over the course of several years with observation intervals {spanning} days, months and years, where the flux variability of 11 T~Tauri stars in the Chamaeleon and Lupus star forming regions was measured at 7 and 15 mm and 3 and 6~cm.
%
Results show that for most sources { are variable to some degree at 7~mm, indicating the presence of emission mechanisms other than thermal dust in some sources.} 
Additionally, { evidence of grain growth to cm-sized} pebbles was found for some sources that also have signs of variable flux at 7~mm.
%
We conclude that multiple processes contributing to the emission are common in T~Tauri stars at 7~mm and beyond, and that a detection {at a} single epoch at radio wavelengths { should not be used to} determine all processes contributing to the emission.

\end{abstract}

\begin{keywords}
protoplanetary discs, star: variables: T~Tauri, radio continuum: planetary systems, radiation mechanisms: thermal, non-thermal 
\end{keywords}


\section{Introduction}

{{ %
Signatures of cm-sized pebbles have been detected in protoplanetary discs by extending the relationship between the dust opacity index, $\beta$, and the 1 to 3~mm spectral slope, $\alpha$, to longer centimetre wavelengths \citep[e.g.][]{2003A&A...403..323T, Rod06,2005ApJ...626L.109W,Lommen09}. %
The results from this relationship ($\beta\sim\alpha-2$) can indicate grain growth {{up to}} $3\lambda$ (where $\lambda$ is the observed wavelength) when $\beta<1$, and {limited to no} grain growth (ISM sized grains) {for $\beta>1$, as long as the grain size $\ll \lambda$} \citep{draine06}.}}

{{
This approach assumes that the long wavelength emission follows the 1 to 3~mm emission, { which is dominated by thermal dust emission} and is constant over time, which is not always the case at longer wavelengths \citep[e.g.][]{Lommen10,Ubach1}.}}

{{
A dozen protoplanetary discs show a ``break" in the spectral slope near 7 mm, {with a shallower spectral slope at longer wavelengths} \citep{Rod06,Lommen09,Lommen10,Ubach1}. This ``break" has been attributed to emission mechanisms related to ionized plasma, adding to the emission from {thermal} dust in these discs. This excess emission at long millimetre wavelengths could be caused by either thermal free-free emission from an ionised wind or by non-thermal processes such as chromospheric emission from the young star, or a combination of both.}}
At 7~mm, both thermal dust emission and thermal free-free emission from an ionised wind have been detected in young stellar objects \citep[e.g.][]{Rod06, Lommen09}, while {{variable emission}} at 3 and 6~cm {{has been used to demonstrate}} a combination of all three emission mechanisms \citep[e.g.][]{Lommen09}.

Temporal monitoring {of the disc emission has been} used to disentangle these three emission mechanisms \citep{Lommen09,Ubach1}. %
{{Flux originating}} from thermal dust emission is { generally assumed to be} constant over time, whereas the flux from thermal free-free and non-thermal emission {{{may be}}} temporally variable. Thermal free-free emission can vary by a factor of $\sim20$--40 per cent over a {{timescale measured in years}} \citep{Skinner:1994fk,2002ApJ...580..459G,Smith:2003qy,Lommen09}, while non-thermal emission mechanisms {{{may}}} vary by a factor of 2 or more on timescales of hours to days \citep{Kuijpers:1985rt,Phillips:1991vn,Phillips:1993yq,Smith:2003qy,Skinner:1994fk,Chiang96}.

Here we present {continuum flux monitoring of T~Tauri stars} at 7 and 15~mm and 3 and 6~cm over timescales of days, months and years. We analyse the data to determine the various contributing emission mechanisms present in {the} discs. We further determine whether it is common for protoplanetary discs to show evidence both of grain growth up to cm-sized pebbles and of multiple emission mechanisms. 


\section{Observations and Data Reduction}
\label{sec2-obs}

This survey targeted 11 T~Tauri~stars in the {{Chamaeleon and Lupus star forming regions -- see Table~\ref{tab-sources}. The sources were selected to overlap with the \citet{Lommen09,Lommen10} and \citet{Ubach1} samples, maximising}} the number of sources {and increasing the monitoring baseline to years} -- see Tables~\ref{tab-obs-log-lupus}~and~\ref{tab-obs-log-cham}.

\begin{table*}
\centering
\caption{List of the 11 sources observed with ATCA in this survey.}  
\small
 \begin{tabular}{llllllc}
	 \hline \hline
	Source	&	RA      & DEC      & Cloud & Distances & T$_{\rm eff}$ &	Wavelengths	\\
			&    (J2000) & (J2000) &           & (pc)           &	 (K)         & (mm)   \\
	\hline \hline									
	 \multicolumn{7}{c}{Chamaeleon}											\\
	\hline						
	CR Cha		&	10 59 06.9	&	-77 01 39.7 & Cha I	& 160	&	4900	&	7	    \\
	CS Cha$^{\Phi}$		&	11 02 24.9	&	-77 33 35.9	& Cha I	& 160 	&	4205	&	7,15   	\\
 	DI Cha$^{\Phi}$		&	11 07 21.6 	&	-77 38 12.0	& Cha I	& 160	&	5860	&	7 	   	\\
	T Cha		&	11 57 13.6 	&	-79 21 31.7	& Isolated &	100	&	5600	&	7,30,60 \\
	Glass I$^{\Phi}$		&	11 08 15.1	&	-77 33 59.0	& Cha I	& 160	&	5630	&	7 	    \\
	SZ Cha		&	10 58 16.7	&	-77 17 17.1	& Cha I	&	160 &	5250	&	7,{ 15}    	\\
	Sz 32		&	11 09 53.4	&	-76 34 25.5	& Cha I	& 160 &	4350	&	7,30,60	\\
	WW Cha		&   11 10 00.1	&	-76 34 57.9	& Cha I	& 160 &	4350	&	7,30,60 	\\
	\hline							
	 \multicolumn{7}{c}{Lupus}											\\
	\hline
	Sz 111		&	16 08 54.7	&	-39 37 43.1	& Lupus 4 & 200 &	3573	&	7,15		\\
	MY Lup		&	16 00 44.6	&	-41 55 29.6	& Lupus 3 &	165 &	5248	&	7,15		\\
	GQ~Lup      &   15 49 12.1	&	-35 39 03.9	& Lupus 1 &  156 &   4300    &   7,15        \\
	\hline	
 \end{tabular}
   	\begin{tablenotes}  
		\item[1] Distances and effective temperatures taken from: (1) \citet{1997A&A...327.1194W}, (2) \citet{1998A&A...330..145V}, (3) \citet{2004ApJ...602..816L}, (4) \citet{1994AJ....108.1071H}, (5) \citet{2008hsf2.book..295C}, (6) \citet{2008A&A...484..281N}, (7) \citet{2012MNRAS.425.2948D}.
		\item[2] $^{\Phi}$ Binary separations of $\leq1^{\prime\prime}$, $4.6^{\prime\prime}$, $2.5^{\prime\prime}$ for CS~Cha, DI~Cha and Glass~I, respectively \citep{2007A&A...467.1147G,1997ApJ...481..378G,1989ApJ...338..262F}.
	\end{tablenotes}
 \label{tab-sources}
 \end{table*}

Continuum observations were conducted with Australia Telescope Compact Array (ATCA) {{during the winter 2012 season. Three compact hybrid array configurations were used, 
providing synthesized beams between 5 to 15 arcsec {at 7~mm}, corresponding to {$\sim800-2400$~au} assuming a distance of 160~pc to the sources.}}
{{The two CABB\footnote{Compact Array Broadband Backend \citep{2011MNRAS.416..832W}.} intermediate frequency bands were divided into}} 2048 channels of 1~MHz width. %

All targets were observed with frequency pairs centred at $43+45$~GHz (7~mm band), five targets\footnote{Only two of the sources have previous 15~mm observations.} at $17+19$~GHz (15~mm band) and three targets at $5.5+9.9$~GHz (6 and 3~cm bands, respectively). %
{{The mm frequency bands setup provided a 4~GHz wide band with an expected RMS of 0.1 and 0.03~mJy for 30~minutes on source for the 7~and~15~mm bands, respectively. The 3 and 6~cm bands were 2~GHz wide with an expected RMS of 0.027 and 0.034~mJy for 30~minutes on source, respectively. In total this survey covered frequencies ranging from 4.5 to 46~GHz.}}

QSO~B1057-79 and QSO~B1600-44 were used as gain calibrators for Chamaeleon and Lupus, respectively.
For all the observations the absolute flux calibration was performed using PKS~1934-638. 
Weather conditions were good throughout the observations. Due to poor \textit{u-v} coverage at the longer baselines {{when using the compact hybrid arrays, and given that fluxes rather than images were the required data products, antenna 6 was not included in the data processing and analysis.}}

The data calibration followed the standard CABB procedure described in the ATCA user guide\footnote{\small{http://www.narrabri.atnf.csiro.au/observing/users\_guide/\\html/atug.html For specific data reduction procedure see \citet{Ubach2014}.}} for all {{wavelengths using}} the software package \texttt{MIRIAD} version 1.5 {{\citep{1995ASPC...77..433S}}}. In order to determine the source fluxes, we used a point source fit. To ensure that this was appropriate, we checked the visibility amplitudes were constant over \textit{u-v} distance for all 11 sources at each epoch. 

For this analysis, the data from each day {at each frequency} was calibrated separately and then the frequency pairs were combined. %
The flux (from the point source fit) and RMS values for all observations were then extracted from the visibilities using {\sc{uvfit}} and {\sc{uvrms}}. 


\section{Results and Discussion}
\label{sec3-results}

{
A summary of the continuum fluxes for the combined frequency pairs are presented in Table~\ref{tab-results}, complemented with fluxes obtained from literature.
The detection rates for the sources observed in this study are as follows: { 9/11} sources were detected at 7~mm, {3/5} at 15~mm\footnote{See Table~\ref{tab-results-extras} for results of sources observed at 15~mm with no previously reported 15~mm fluxes.}, and {0/3} sources at both 3~cm and 6~cm.
A $3\sigma$ upper limit is provided for non-detections. 

{ We present in Figure~\ref{fig-epochs} both the spectrum, or radio flux as a function of wavelength (left), and the 7~mm flux versus time (right) for the 11 sources in our survey. Detections are denoted by a star symbol and upper limits by an arrow. Error bars include the uncertainty in the flux fit, as well as the primary flux calibration uncertainties of 20 per cent for 1.2~mm, 30 per cent for 3~mm, and 10 per cent for 7~mm, 15~mm, 3~cm and 6~cm. In the spectrum plots, the red solid line represents the estimated emission due to thermal dust, which was fit using the least squares method between the 1~mm and 3~mm fluxes and extrapolating this spectral slope, $\alpha$, to longer wavelengths. Fluxes at 1 and 3~mm were taken from \citet{Ubach1} and references therein. The estimated 7~mm thermal dust component, $F_{\rm 7mm}{\rm (therm)}$, is denoted by a green diamond. The grey shaded area represents the region of uncertainty in the spectral slope using standard quadratic error propagation.} 
{ If we take a `slice' through the radio spectrum at 7~mm, and plot each observed 44~GHz flux against time (as presented in Table~\ref{tab-results}), this gives us the temporal monitoring plots on the right. Here the grey region is the same as in the spectrum plot, and represents the uncertainty of the extrapolation of the 7~mm thermal flux estimate, $F_{\rm 7mm}{\rm (therm)}$, which is now represented by a green line.  We will use both of these plots to help understand the emission processes in each system.}
%

 \begin{table*}
 \caption{Summary of the temporal monitoring results at 7 and 15~mm and 3 and 6~cm for our sample of 11 T~Tauri stars. (1) Source name. (2) Julian date of the observation. (3)  Time from previous observation in days. (4) Combined frequency. (5) Point flux (3$\sigma$ upper limit for non-detections). (6) RMS.  (7) Reference. {See Table~\ref{tab-results-extras} for new 15~mm observations of sources with no previously reported 15~mm fluxes.}}
 \centering
 \footnotesize{
 \begin{tabular}{lcccccc}
 \hline \hline
 (1) & (2) & (3) & (4) & (5) & (6) & (7) \\
 Source	&		Julian 	&	Time from	 &		Freq. 	&	Flux & RMS &			Reference \\
 & date & prev. obs. (days) & (GHz) & (mJy) & (mJy/Beam) & \\
 \hline\hline
CR~Cha	&	2454980	&	---	&		44		&		$1.7\pm0.1$		&		0.1		&	\citet{Ubach1}	\\
	&	2456143	&	1163	&		44		&		$1.3\pm0.2$		&		0.1		&	This work	\\
\hline													
CS Cha	&	2454583	&	---	&	44	&	$1.00\pm0.28$	&	0.129	&	\citet{Lommen09}	\\
	&	2454677	&	94	&	44	&	$<0.71$	&	0.238	&	\citet{Lommen09}	\\
	&	2454678	&	1	&	44	&	$1.38\pm0.26$	&	0.218	&	\citet{Lommen09}	\\
	&	2454980	&	302	&		44		&		$1.5\pm0.1$		&		0.1		&	\citet{Ubach1}	\\
	&	2456143	&	1163	&		44		&		$1.5\pm0.2$		&		0.1		&	This work	\\
	&	2456223	&	81	&		44		&		$0.6\pm0.1$		&		0.3		&	This work	\\
	&	2455753	&	---	&		18		&		$<0.3$		&		0.1		&	\citet{Ubach1}	\\
	&	2456131	&	378	&		18		&		$0.20\pm0.02$		&		0.03		&	This work	\\
\hline													
DI~Cha	&	2454980	&	---	&	44	&	$0.9\pm0.1$	&	0.1	&	\citet{Ubach1}	\\
	&	2456143	&	1163	&	44	&	$<0.2$	&	0.1	&	This work	\\
\hline													
T~Cha	&	2454980	&	---	&	44	&	$3.0\pm0.1$	&	0.1	&	\citet{Ubach1}	\\
	&	2456143	&	1163	&	44	&	$4.1\pm0.2$	&	0.1	&	This work	\\
	&	2455761	&	---	&	9.9	&	$<0.3$	&	0.1	&	\citet{Ubach1}	\\
	&	2456123	&	363	&	9.9	&	$<0.15$	&	0.05	&	This work	\\
	&	2455761	&	---	&	5.5	&	$0.3\pm0.1$	&	0.1	&	\citet{Ubach1}	\\
	&	2456123	&	363	&	5.5	&	$<0.2$	&	0.08	&	This work	\\
\hline													
Glass I	&	2454980	&	---	&	44	&	$0.6\pm0.2$	&	0.1	&	\citet{Ubach1}	\\
	&	2456143	&	1163	&	44	&	$<0.3$	&	0.1	&	This work	\\
\hline													
SZ Cha	&	2454980	&	---	&	44	&	$0.7\pm0.1$	&	0.1	&	\citet{Ubach1}	\\
	&	2456143	&	1163	&	44	&	$0.6\pm0.1$	&	0.1	&	This work	\\
	&	2456223	&	81	&	44	&	$<0.3$	&	0.1	&	This work	\\
\hline													
Sz~32	&	2454558	&	---	&	44	&	$0.8\pm0.1$	&	0.2	&	\citet{Lommen09}	\\
	&	2454980	&	422	&	44	&	$1.2\pm0.1$	&	0.2	&	\citet{Ubach1}	\\
	&	2456143	&	1163	&	44	&	$0.5\pm0.2$	&	0.1	&	This work	\\
	&	2455761	&	---	&	9.9	&	$<0.3$	&	0.1	&	\citet{Ubach1}	\\
	&	2456123	&	363	&	9.9	&	$<0.15$	&	0.05	&	This work	\\
	&	2455761	&	---	&	5.5	&	$<0.3$	&	0.1	&	\citet{Ubach1}	\\
	&	2456123	&	363	&	5.5	&	$<0.1$	&	0.04	&	This work	\\
	\hline													
WW~Cha	&	2454980	&	---	&	44	&	$9.1\pm0.35$	&	0.2	&	\citet{Ubach1}	\\
	&	2456143	&	1163	&	44	&	$3.9\pm0.2$	&	0.1	&	This work	\\
	&	2455761	&	---	&	9.9	&	$<0.3$	&	0.1	&	\citet{Ubach1}	\\
	&	2456123	&	363	&	9.9	&	$<0.15$	&	0.05	&	This work	\\
	&	2455761	&	---	&	5.5	&	$<0.45$	&	0.1	&	\citet{Ubach1}	\\
	&	2456123	&	363	&	5.5	&	$<0.12$	&	0.04	&	This work	\\
\hline													
Sz 111	&	2454680	&	---	&	44	&	$<0.6$	&	0.2	&	\citet{Lommen10}	\\
	&	2454980	&	300	&	44	&	$0.5\pm0.1$	&	0.1	&	\citet{Ubach1}	\\
	&	2456143	&	1163	& 	44	& 	$0.6\pm0.2$	& 	0.1	& 	This work	\\
	&	2456223	&	81	&	44	&	$0.4\pm0.1$	&	0.03	&	This work	\\
\hline													
MY Lup	&	2454680	&	---	&	45	&	$1.3$	&	0.1	&	\citet{Lommen10}	\\
	&	2454980	&	300	&	44	&	$1.1\pm0.1$	&	0.1	&	\citet{Ubach1}	\\
	&	2456143	&	1163	& 	44	& 	$1.0\pm0.2$	& 	0.1	& 	This work	\\
	&	2456223	&	81	&	44	&	$1.2\pm0.2$	&	0.03	&	This work	\\
\hline												
GQ Lup	&	2455430	&	---	&	44	&	$0.6\pm0.1$	&	0.04	&	\citet{Ubach1}	\\
	& 	2455753	& 	323	& 	44	& 	$0.6\pm0.3$	&	0.07	&	\citet{Ubach1}	\\
	&	2456144	&	391	&	44	&	$1.2\pm0.1$	&	0.05	&	This work	\\
	&	2456223	&	79	&	44	&	$0.5\pm0.1$	&	0.04	&	This work	\\
	&	2455751	&	---	&	18	&	$0.07\pm0.03$	&	0.01	&	\citet{Ubach2014}	\\
	& 	2455761	& 	10	& 	18	& 	$0.08\pm0.04$	&	0.02	&	\citet{Ubach2014}	\\
	&	2456223	&	462	&	18	&	$<0.09$	&	0.03	&	This work	\\
	\hline
 \end{tabular}
  }
  \label{tab-results}		
\end{table*}

 \subsection{Temporal monitoring}
 \label{sub-sec-temp}

{  The aim of our temporal monitoring campaign is to try and {{disentangle}} the different physical {{processes}} contributing to the radio emission in this sample of T~Tauri stars. While a number of millimetre and centimetre surveys of YSOs have been conducted {{\citep[e.g.][]{H1993,Andrews07}}}, few monitoring {{surveys}} have been carried out due to their intense resources requirement.  Given that YSOs are inherently variable {{\citep[e.g.][]{1994AJ....108.1906H}}}, flux monitoring on timescales of a day, weeks and years are required to understand the contributing processes to their emission. Here we compare fluxes obtained from a few days apart to 100s of days apart, as presented in Table~\ref{tab-results}.
}
 

{ From our previous observations \citep{Ubach1}, our sample could be divided into two distinct groups: (i) those with a constant spectral slope from 1-7~mm (CS~Cha, Glass~I, SZ~Cha, WW~Cha, MY~Lup, Sz~111, GQ~Lup), and (ii) those with a ``break'' in the mm spectral slope at 7~mm (CR~Cha, DI~Cha, T~Cha and Sz~32). This ``break'' 
is an indication that other forms of emission at 7~mm are present in addition to thermal dust emission. 
In \citet{Ubach1} a least squares fit was used to determine the spectral slope between 1 and 3~mm, $\alpha$, and extrapolated to  7~mm to estimate the thermal dust component $F_{\rm 7mm}{\rm(therm)}$. Table~\ref{tab-results2} presents the estimated thermal dust emission at 7~mm, the maximum observed 7~mm flux, $F_{\rm 7mm}{\rm(max)}$, and the fractional 7~mm excess, given by $(F_{\rm 7mm}{\rm (max)} - F_{\rm 7mm}{\rm(therm)}) / F_{\rm 7mm}{\rm(therm)}$, for this sample of 11 protoplanetary discs. 
} 

{ Using a combination of the spectrum and the temporal monitoring plots in Figure~\ref{fig-epochs}, along with the data in Table~\ref{tab-results2}, we can start to understand the degree and potential source of variability in each system\footnote{Note that the fractional 7~mm excess presented in Table~\ref{tab-results2} does not take into consideration the errors in the estimated 7~mm thermal dust emission, and so can only be considered a very crude measure of any emission excess above thermal dust.}.
If we start with the (crude) fractional 7~mm excess, we can place the 11 sources into three groups: those with high to extreme fractional excesses above 1.5; those with a moderate fractional excess between 0.5--1.0; and those with a small fractional excess less than 0.5.
}

{
Of the high to extreme 7~mm fractional excess sources, we can see that CR~Cha and DI~Cha have excesses of order 2 above expected thermal dust, while T~Cha and Sz~32 have much larger excesses of order 4--5. Looking at the radio spectrum of the two most extreme sources, we can clearly see a break in the spectral slope at 7~mm, excess emission and flux variability in T~Cha and Sz~32. Their 7~mm temporal monitoring plots show fluxes well in excess of that estimated for thermal dust, i.e. the observed fluxes are outside of the grey regions of the plots.  Over the 3 year period of the observations, the 7~mm fluxes of T~Cha and Sz~32 varied by a factor of $\sim 0.4$ and $1.4$ respectively. The 3 and 6~cm flux monitoring of both sources also suggests variably over the course of a year by 30--50 per cent, and detailed analysis of the 15~mm and $3+6$~cm intra-epoch fluxes by \citet{Ubach1} show variability of a factor of 2 on the timescale of tens of minutes. 
There is also a break in the spectral slope at 7~mm seen in CR~Cha and DI~Cha, and over the 3 year period of 7~mm observations, both show variably of a factor of 0.3 and 3 respectively. They have only a single observation at 15~mm and intra-epoch analysis by \citet{Ubach1} of CS~Cha shows 15~mm variability of a factor of 2 on timescales of tens of minutes.
For all four sources it seems likely that a mix of thermal dust, thermal free-free emission and non-thermal processes contribute to the radio emission.
}

{ 
The sources with moderate fractional excess include WW~Cha, and GQ~Lup. None of these sources have a break in their spectral slope at 7~mm, and their 7~mm fluxes all lie within the region of uncertainty of the extrapolated 7~mm thermal dust emission. 
Three of the four 7~mm fluxes recorded for GQ~Lup over a 2.2 year period are consistent, though there is variability of a factor of 2.4 in just under 3 months.  GQ~Lup was also observed at 15~mm over a period of almost 1.3 years, and all three fluxes are consistent within the flux fit errors -- see Figure~\ref{fig-epochs-gqlup}. While not exactly simultaneous, the two 7~mm and 15~mm observations that were made about 470 days apart give consistent (non-varying) fluxes in both bands. }

{
There have been many observations of WW Cha -- see Table~\ref{tab-results-wwcha} for a summary.  In Table~\ref{tab-results} we present just the 44, 18, 9.9 and 5.5~GHz observations for the  sources in our survey,{while as it can be seen in} Table~\ref{tab-results-wwcha}, WW~Cha has been observed at various wavelengths in the 7~mm, 15~mm and 3 \& 6~cm bands. In the spectrum of WW~Cha in Figure~\ref{fig-epochs} we present all the data from Table~\ref{tab-results-wwcha}, with the red inverted triangles representing detections and upper limits obtained by \citet{Lommen09}. The 7~mm monitoring data shows only the 44~GHz data for consistency. The cm-band data clearly shows a lot of variability, by factors of 2--4, and the two 7~mm fluxes taken 3.2 years apart vary by a factor of 2.3.
With the radio monitoring data to date, it appears that Sz~111 and GQ~Lup are dominated by thermal dust emission, though there is variability at 7~mm in GQ~Lup. WW~Cha likely has a mix of thermal dust, thermal free-free emission and non-thermal processes contributing to the radio emission.
}

{ 
The sources with low fractional excess include CS~Cha and Glass~I. Once again, none show a break in their spectral slope at 7~mm, and their 7~mm fluxes all lie within the region of uncertainty of the extrapolated 7~mm thermal dust estimate. Glass~I vary by a factor of 2.3 and 2 over 3.2 years. CS~Cha varies by about a factor of 3 over 3 months. This magnitude of variability for Glass~I and CS~Cha suggests that the non-thermal emission as well as thermal dust contribute to the overall emission. 
}

{ 
Finally, we find that just three sources in our sample, SZ~Cha, Sz 111 and MY~Lup, have no observed variability at 7~mm over a timescale of 100s of days, suggesting that the detected continuum flux emission is due to thermal dust emission, whereas all other sources exhibit flux variability on the time scale of months to years of at least 30 per cent and up to a factor of 3 at 7~mm is observed, which provides evidence of the presence of excess emission from mechanisms other than thermal dust emission. However, the 7~mm fluxes for these three sources are also lower than expected from spectral fit. This perhaps is indicative that the calculated $\alpha$ is too shallow. This appears to be supported by the lower 15~mm fluxes in Sz~111 and MY~Lup.} 
{{This survey highlights the significance of flux monitoring over multiple timescales, for one flux detection will not always be a good indication of a pure thermal dust emission source at 7~mm and beyond.  
}}

\begin{table*}
\centering
\caption{Summary of 7~mm analysis of the spectral slope, estimated 7~mm thermal dust emission, maximum 7~mm emission observed, and the fractional 7~mm excess above thermal.  The $\alpha$ values are from \citet{Ubach1}, except for WW~Cha which was obtained from \citet{Lommen09}. The estimated thermal dust is obtained by extrapolating $\alpha$ to  44~GHz. The fractional 7~mm excess above thermal is the estimated 7~mm excess, $(F_{\rm 7mm}{\rm (max)}- F_{\rm 7mm}{\rm (therm)})$, divided by the estimated thermal dust, $F_{\rm 7mm}{\rm (therm)}$.} 
\begin{tabular}{ccccc}
	\hline \hline	
		Source	&	$\alpha$	&	F$_{\rm 7mm}$(therm)		&	F$_{\rm 7mm}$(max)	&Fractional 		\\
		 &  & (mJy) & (mJy) & 7mm excess \\
	\hline\hline
    CR~Cha	&	3.4	&	0.6	&	1.7	&	1.8	\\
    CS~Cha	&	2.9	&	1.2	&	1.5	&	0.3	\\
    DI~Cha	&	3.2	&	0.3	&	0.9	&	2.0	\\
    T~Cha	&	3.1	&	0.8	&	4.1	&	4.1	\\
    Glass~I	&	3.1	&	0.5	&	0.6	&	0.2	\\
    SZ~Cha	&	2.9	&	0.8	&	0.7	&	-0.1	\\
    Sz~32	&	3.8	&	0.2	&	1.2	&	5.0	\\
    WW~Cha	&	2.8	&	5.5	&	9.1	&	0.7	\\
    Sz~111	&	2.5	&	1.1	&	0.6	&	-0.5	\\
    MY~Lup	&	2.3	&	1.8	&	1.2	&	-0.3	\\
    GQ~Lup	&	2.2	&	0.7	&	1.2	&	0.7	\\
	\hline
\end{tabular}
\label{tab-results2}
\end{table*}

\begin{figure*}
\centering
      {\includegraphics[width=0.80\columnwidth]{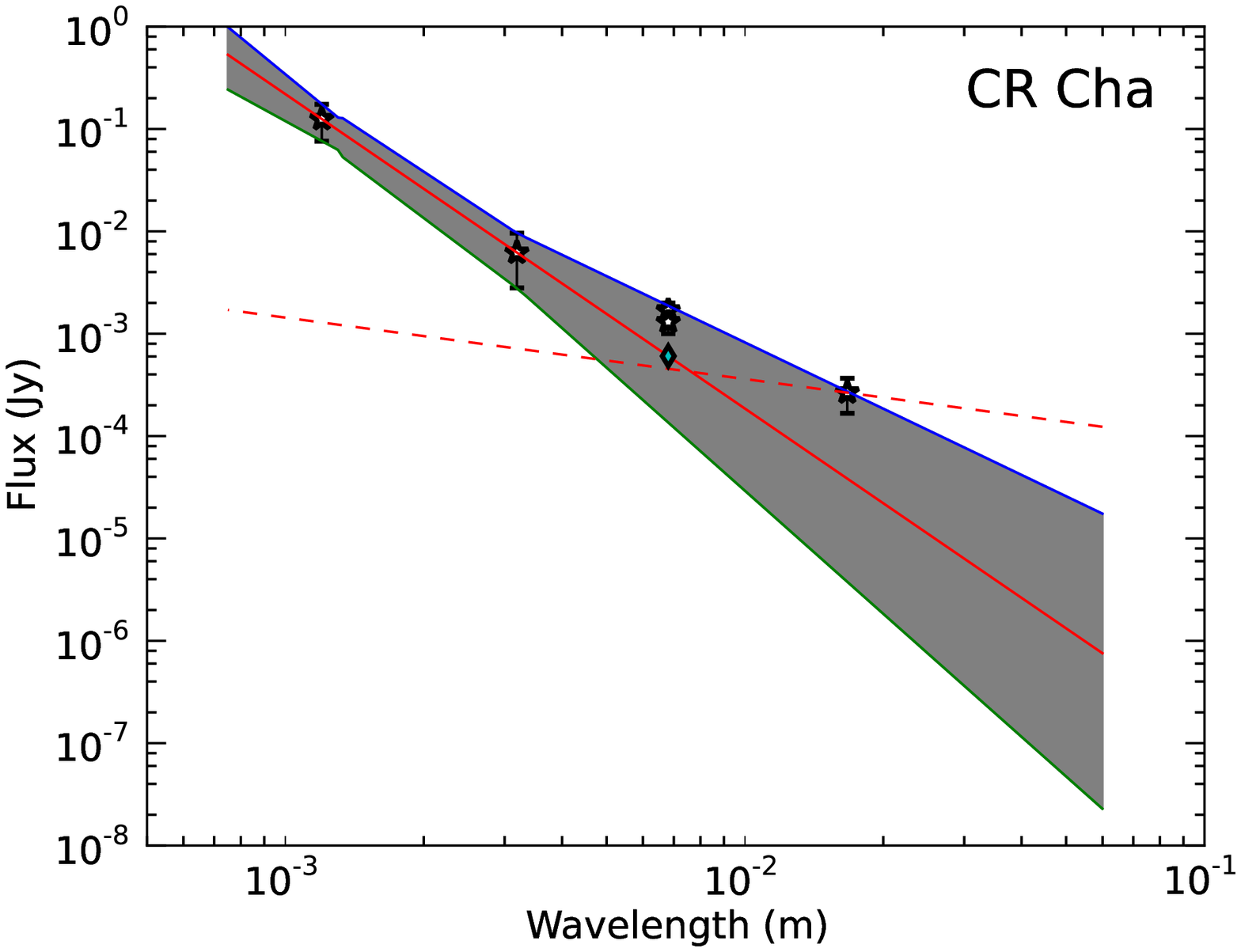}}\quad
      {\includegraphics[width=0.80\columnwidth]{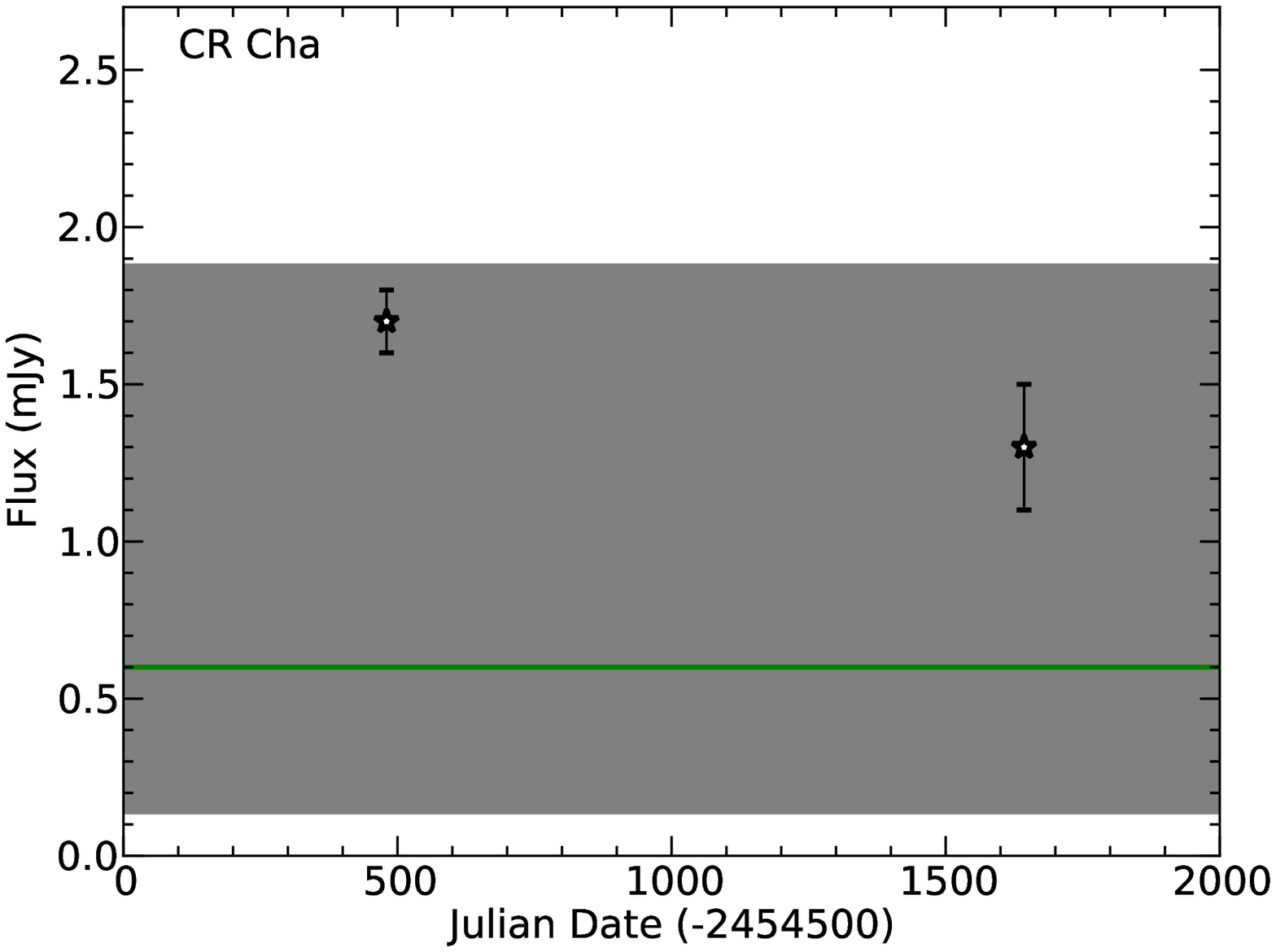}}\quad
      {\includegraphics[width= 0.80\columnwidth]{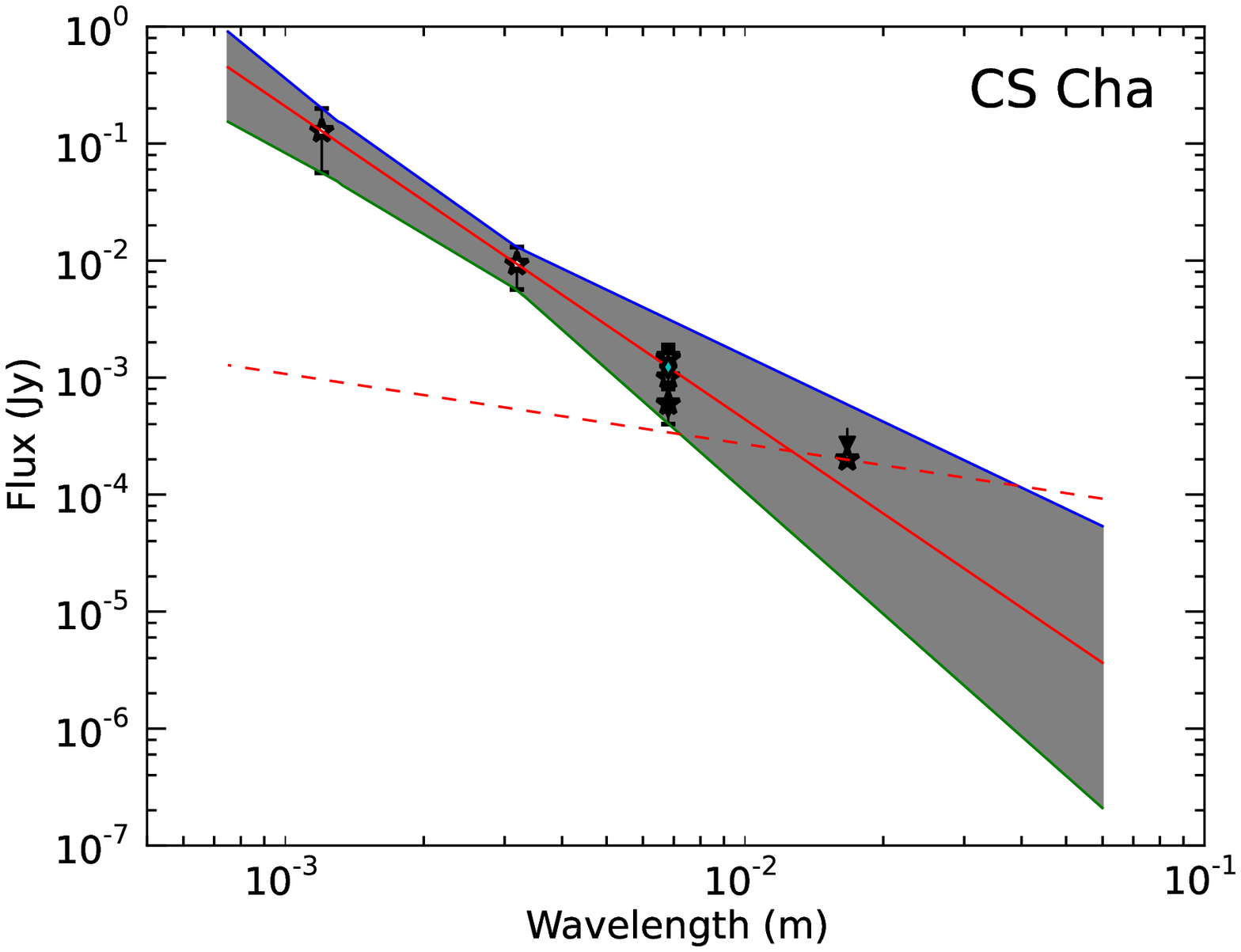}}\quad
      {\includegraphics[width=0.80\columnwidth]{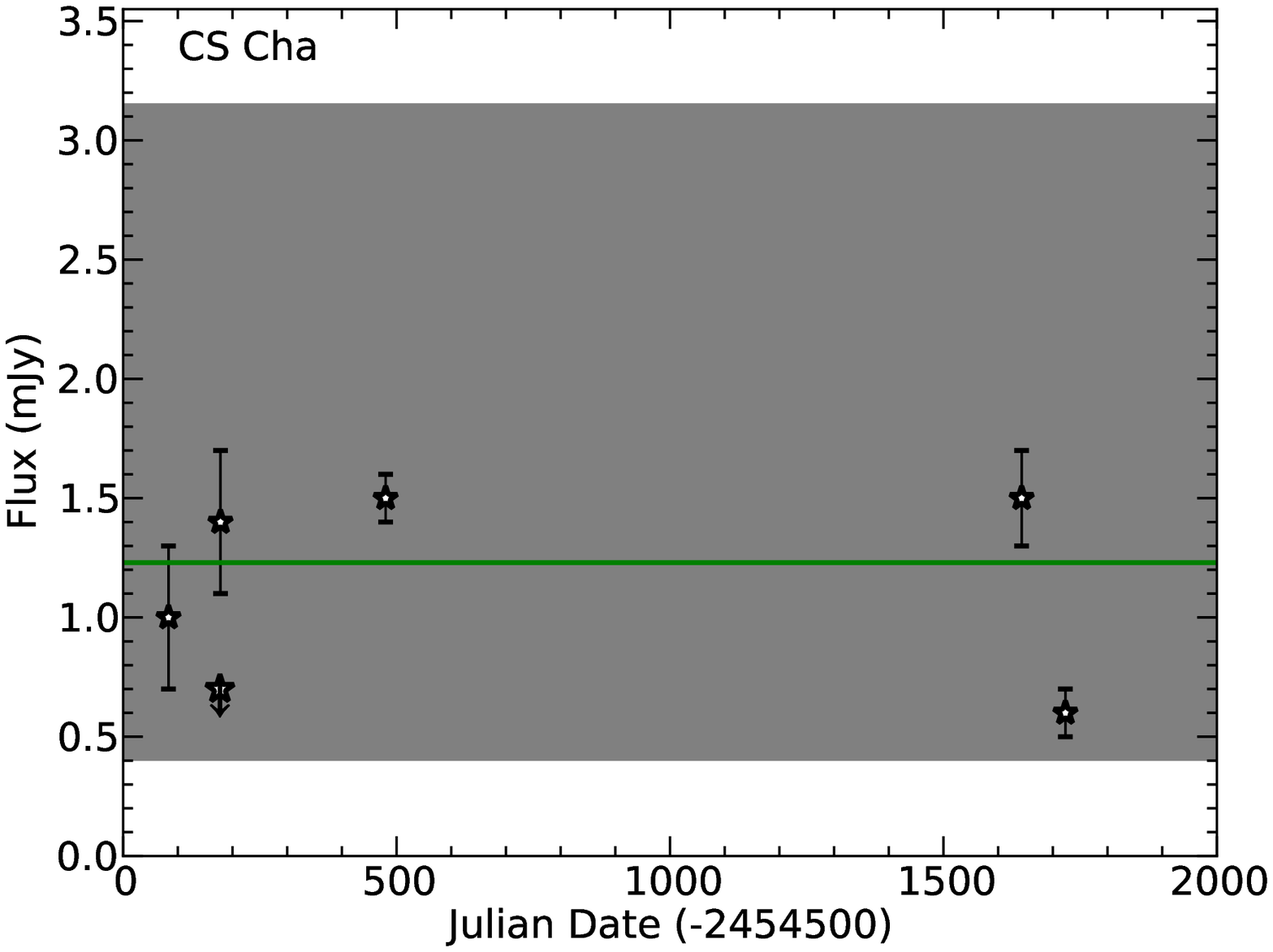}}\quad
      {\includegraphics[width= 0.80\columnwidth]{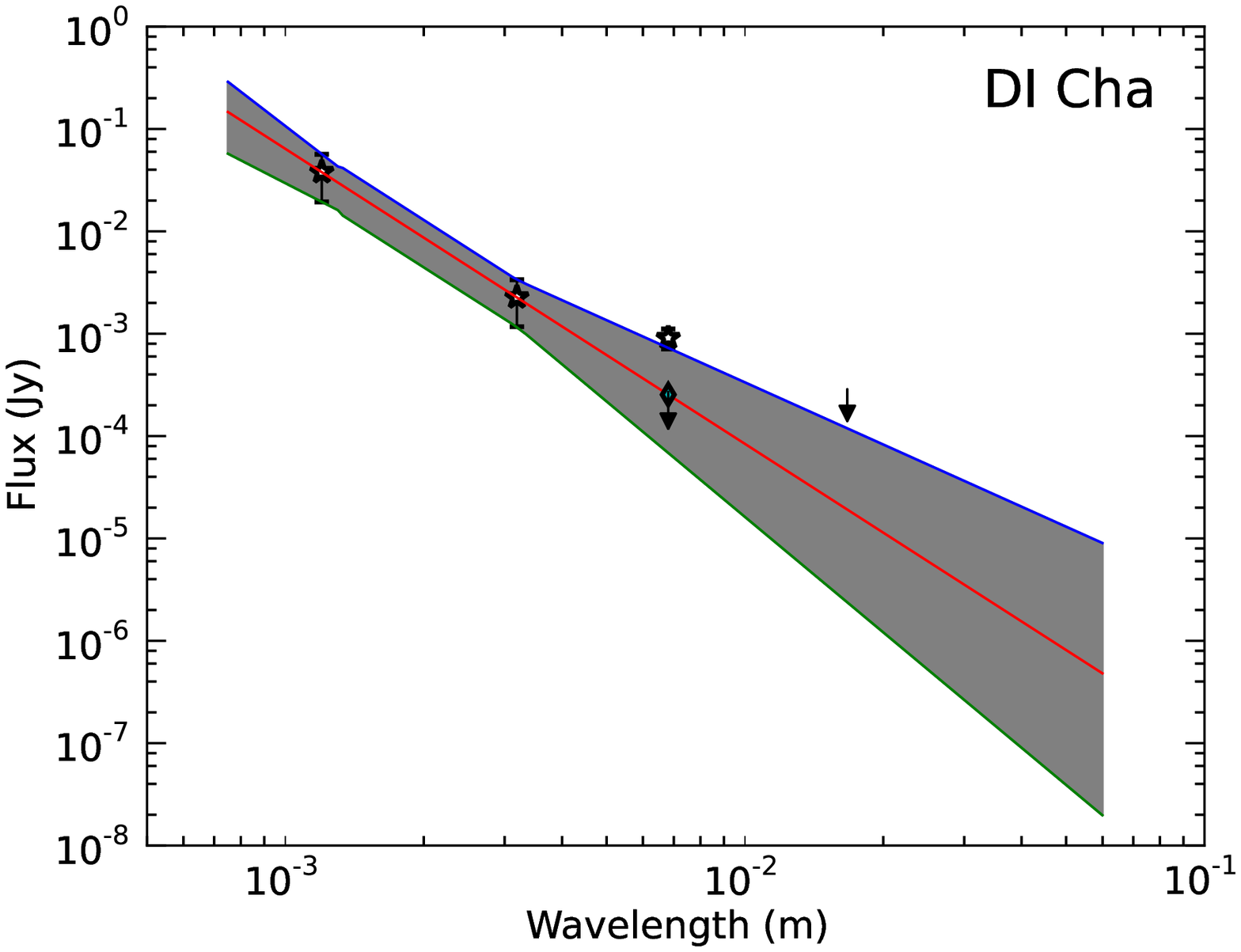}}\quad
      {\includegraphics[width=0.80\columnwidth]{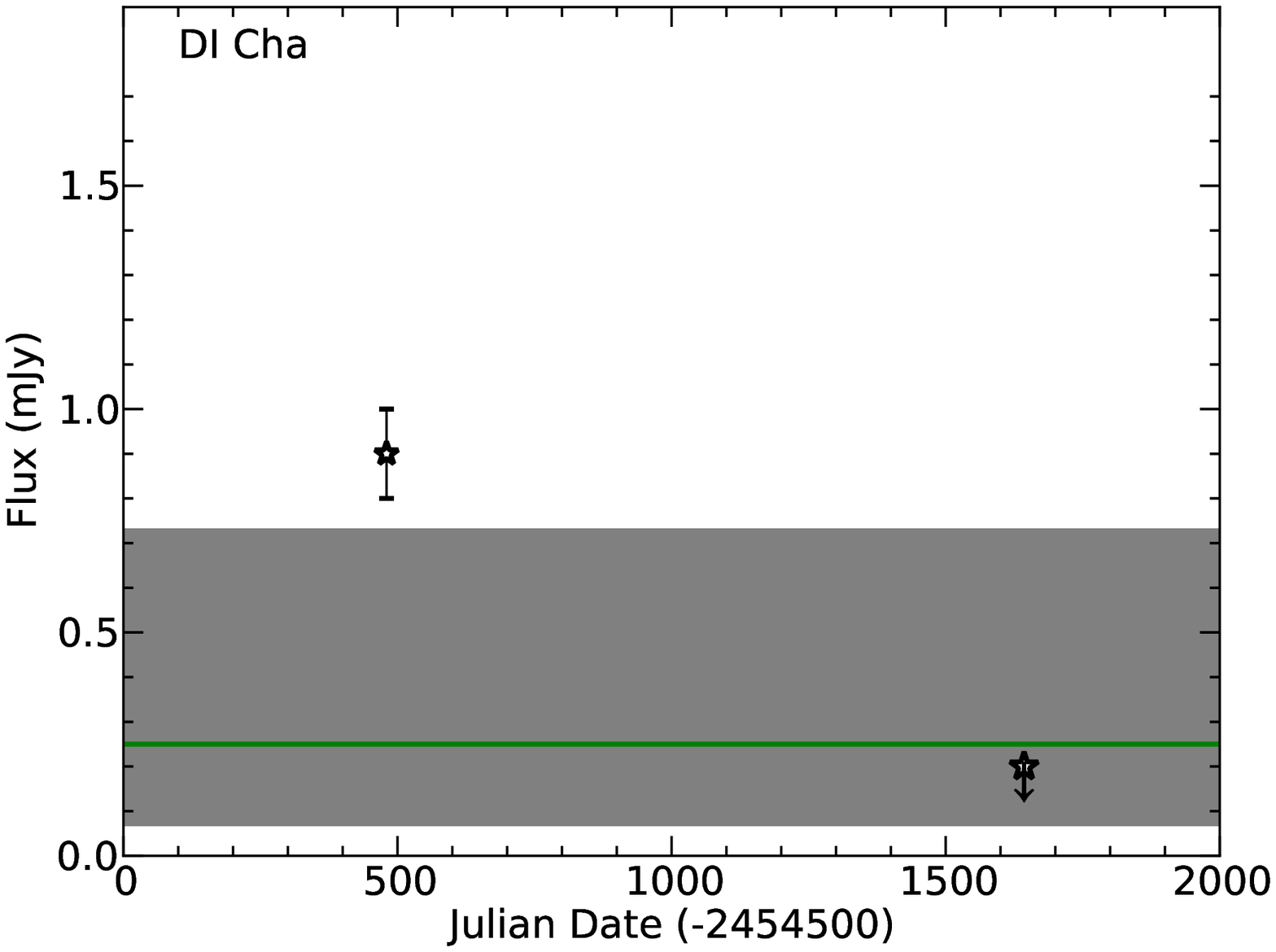}}\quad      
      {\includegraphics[width= 0.80\columnwidth]{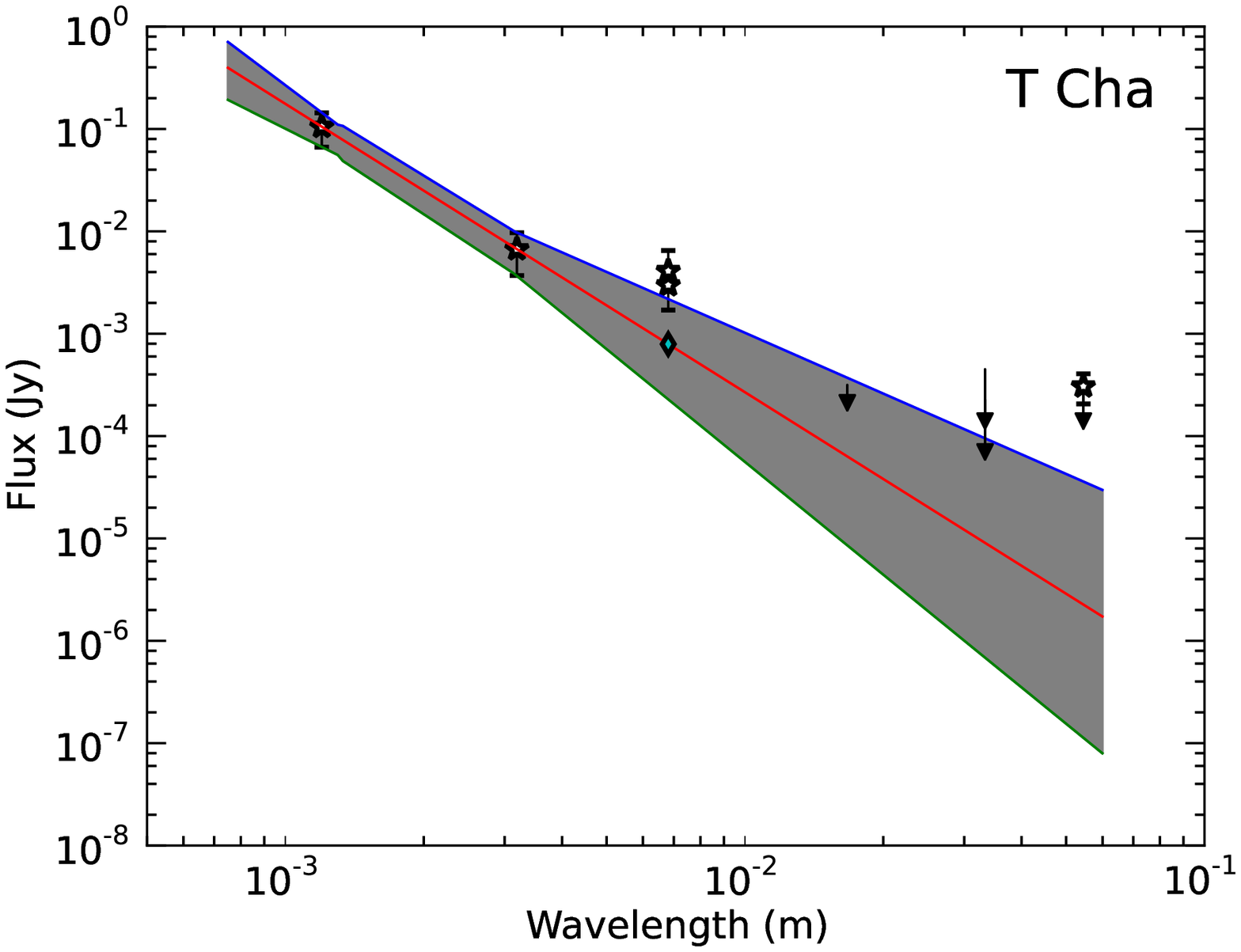}}\quad
      {\includegraphics[width=0.80\columnwidth]{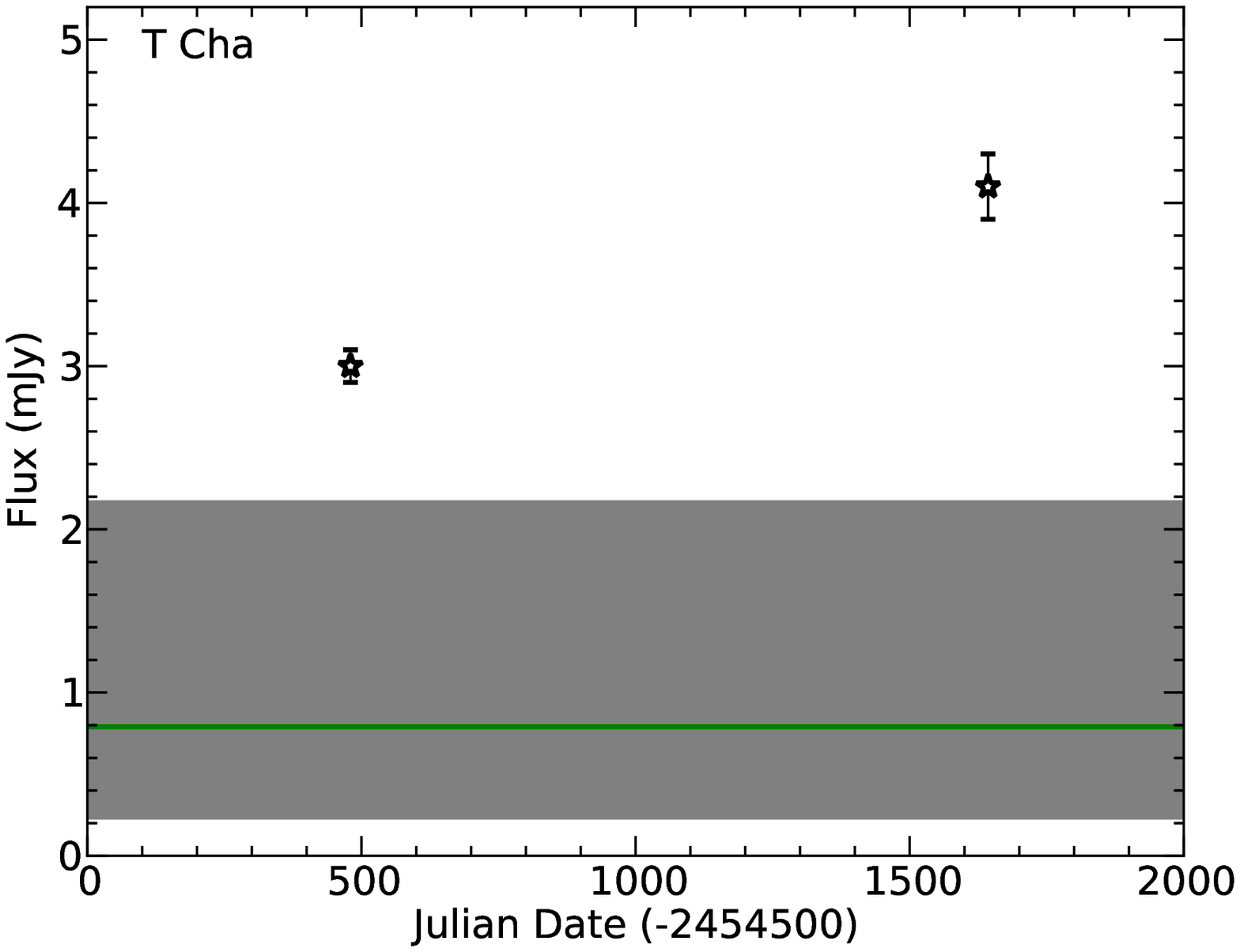}}
\caption[\label{fig-epochs}]{Radio spectrum (left) and temporal flux monitoring (right) for each source in the survey.  Detections are denoted by a star and upper limits by an arrow. Errors bars include the flux fit uncertainty and primary flux calibration uncertainties. The red solid line in the spectrum (left) represents the 1--3~mm spectral slope $\alpha$, the red dashed line the expected $\alpha_{\rm ff} = -0.6$, and the grey shaded area represents the region of uncertainly in the estimated thermal dust emission from standard error propagation in the spectral slope calculation. The green diamond (left) and green line (right) correspond to the estimated thermal dust emission at 7~mm. Continued on the next page.}
\end{figure*}

\begin{figure*}	
\ContinuedFloat\centering 
      {\includegraphics[width= 0.80\columnwidth]{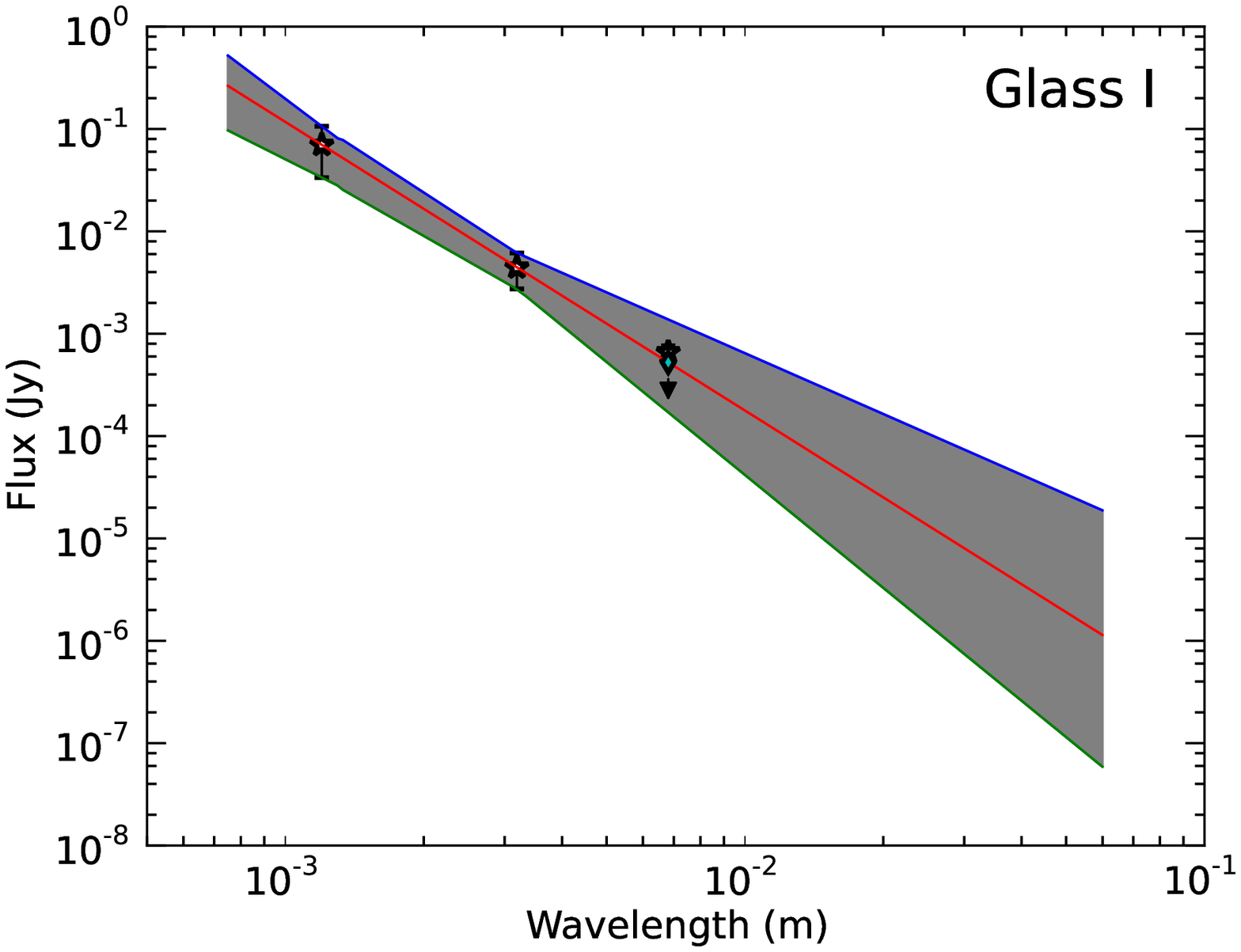}}\quad
      {\includegraphics[width=0.80\columnwidth]{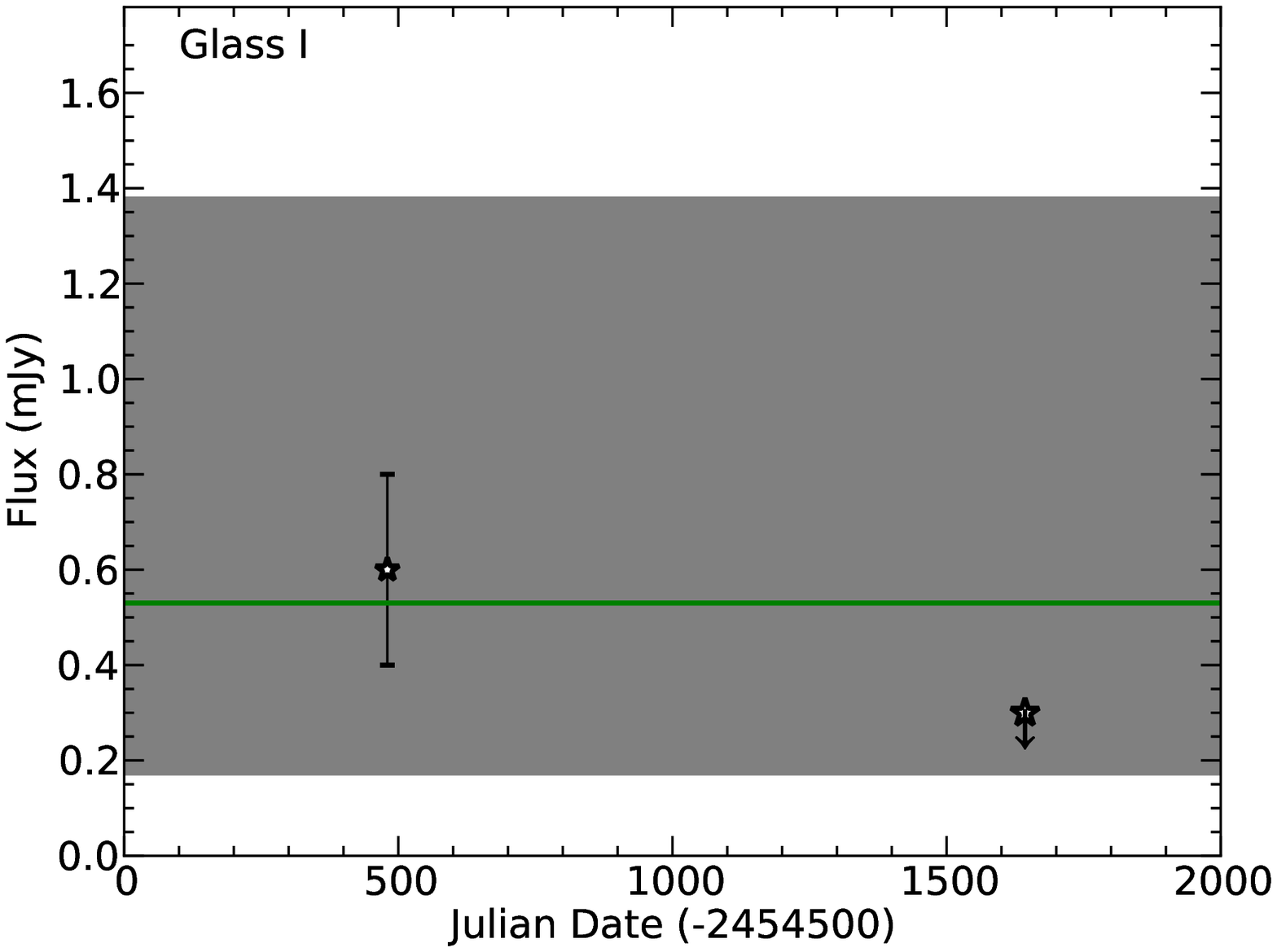}}\quad
      {\includegraphics[width= 0.80\columnwidth]{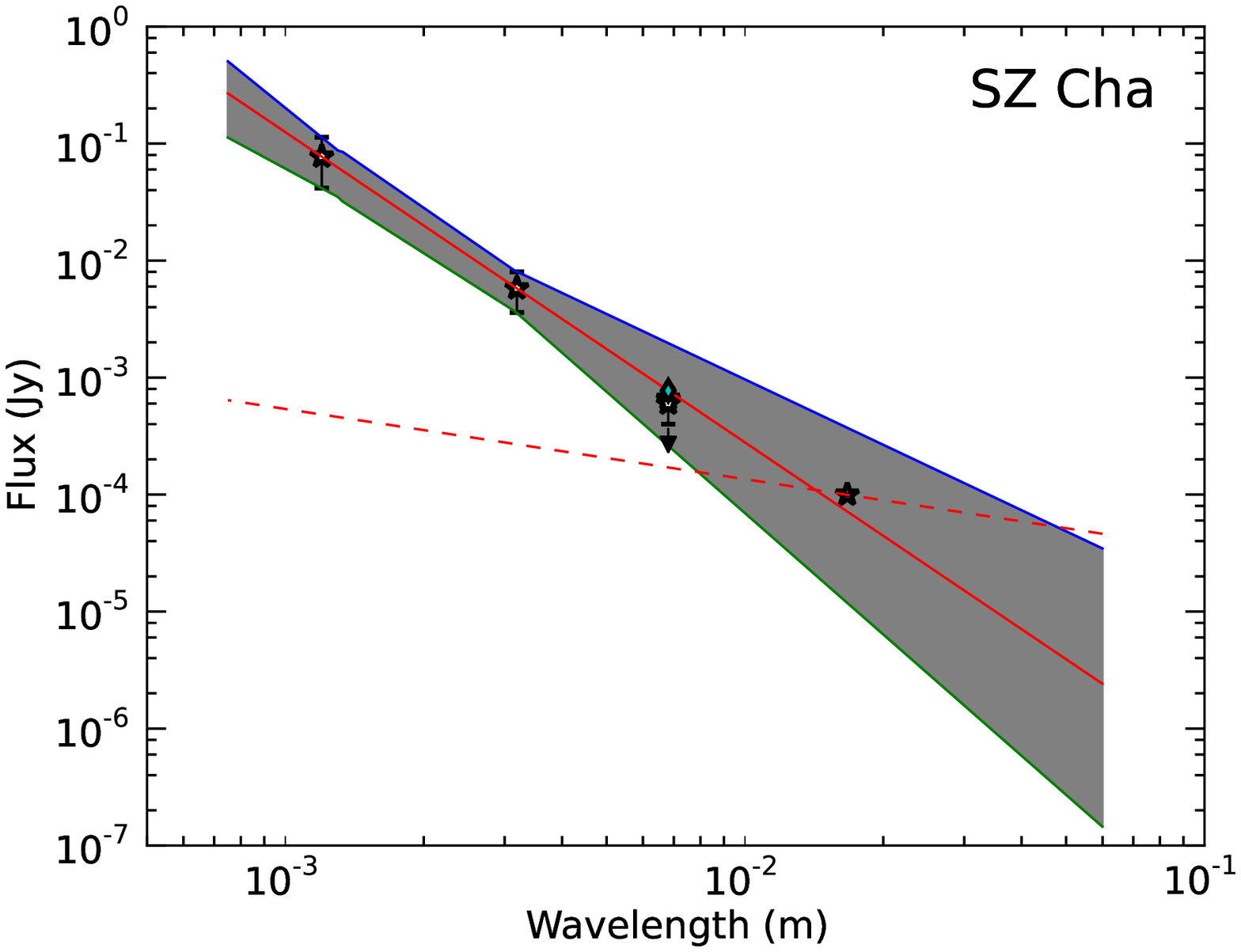}}\quad
      {\includegraphics[width=0.80\columnwidth]{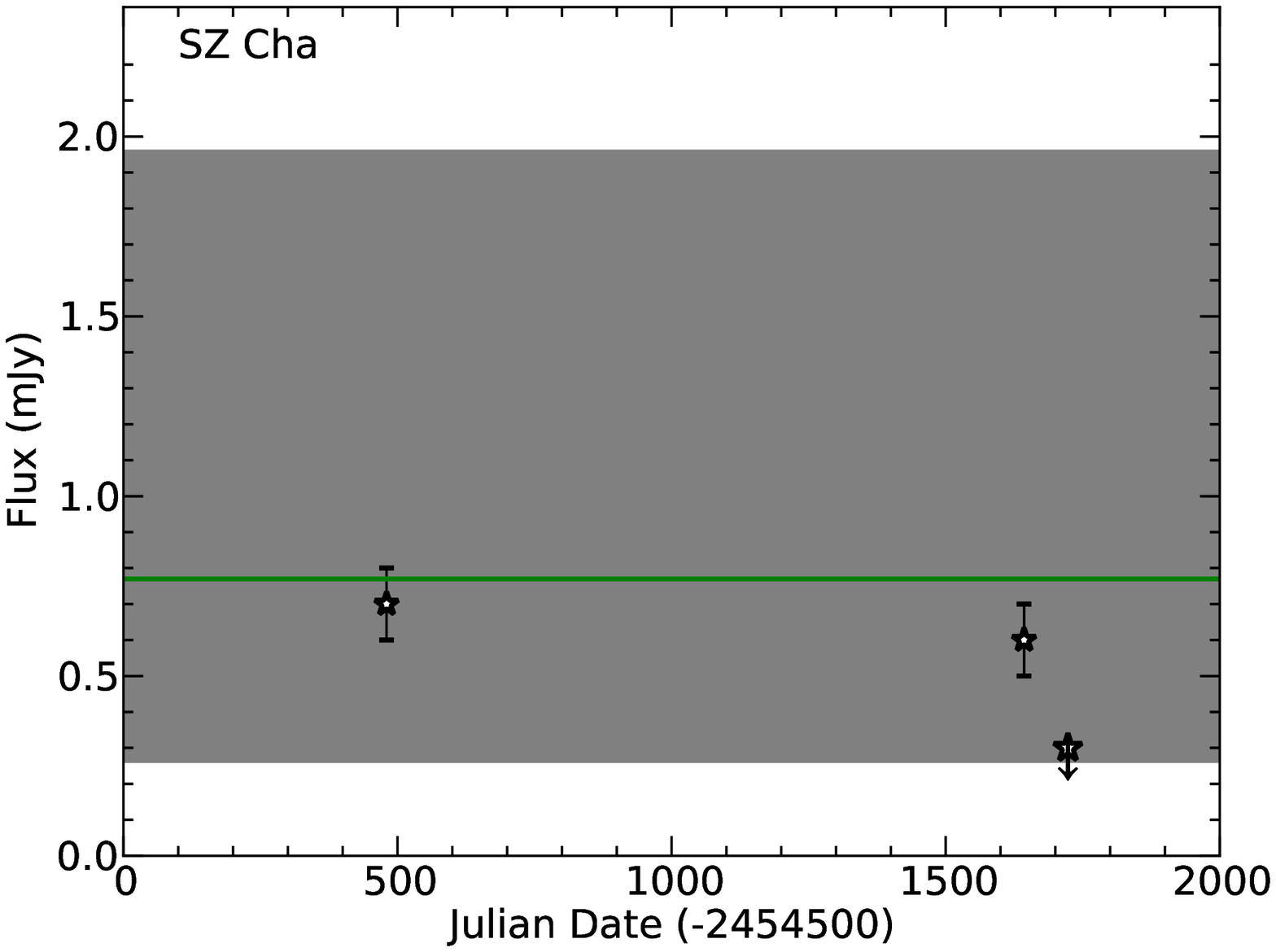}}\quad
      {\includegraphics[width= 0.80\columnwidth]{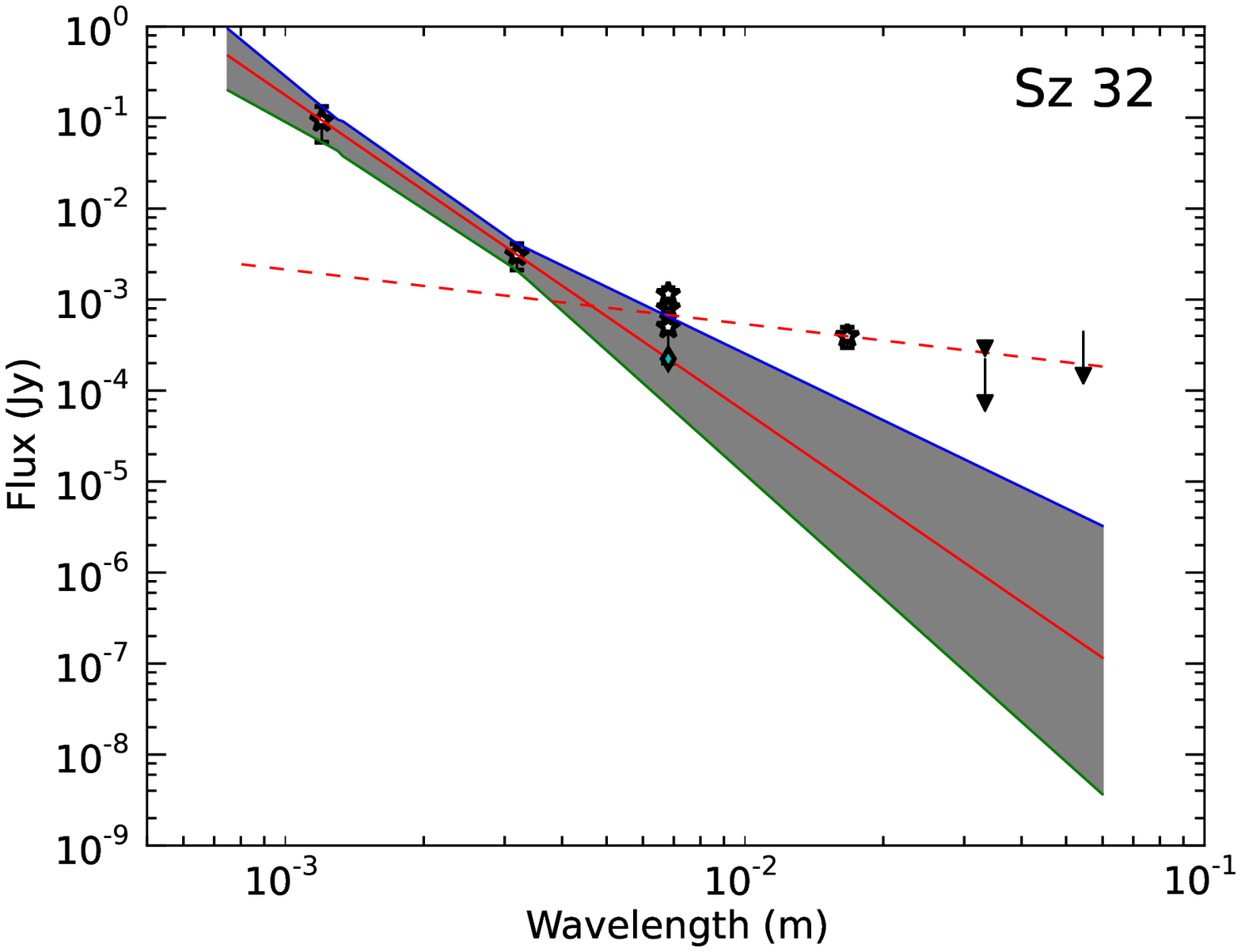}}\quad
      {\includegraphics[width=0.80\columnwidth]{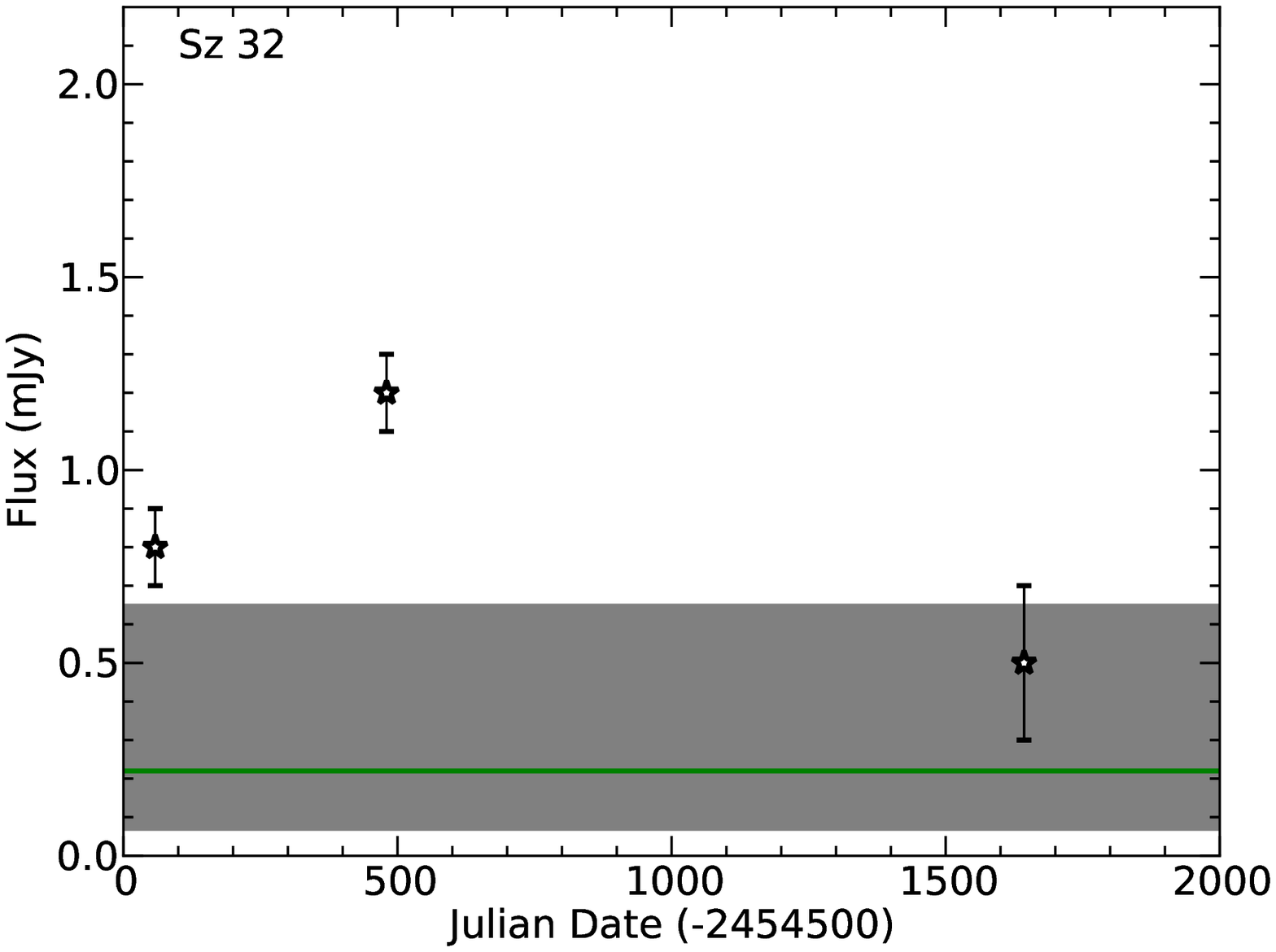}}\quad
      {\includegraphics[width= 0.80\columnwidth]{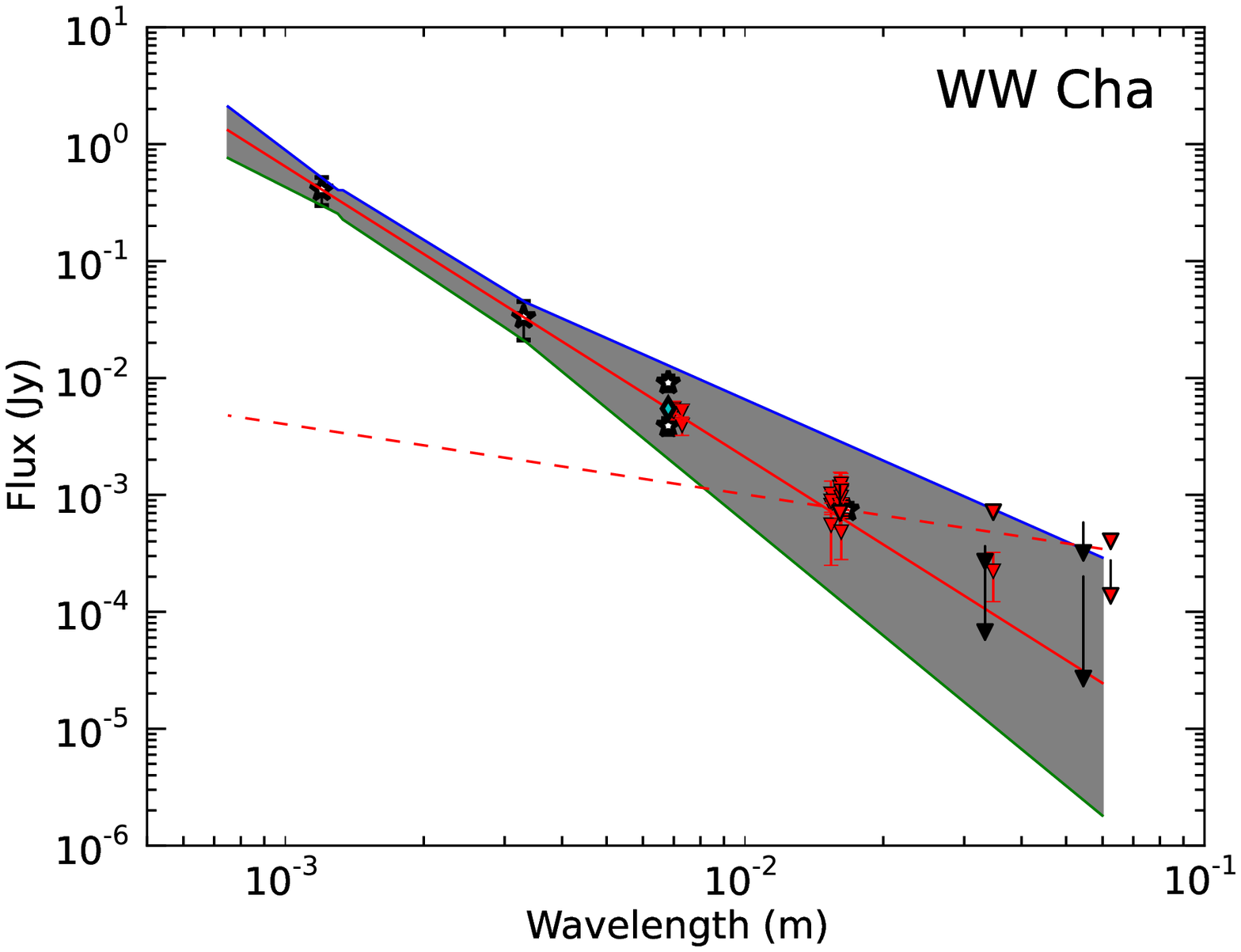}}\quad
      {\includegraphics[width=0.80\columnwidth]{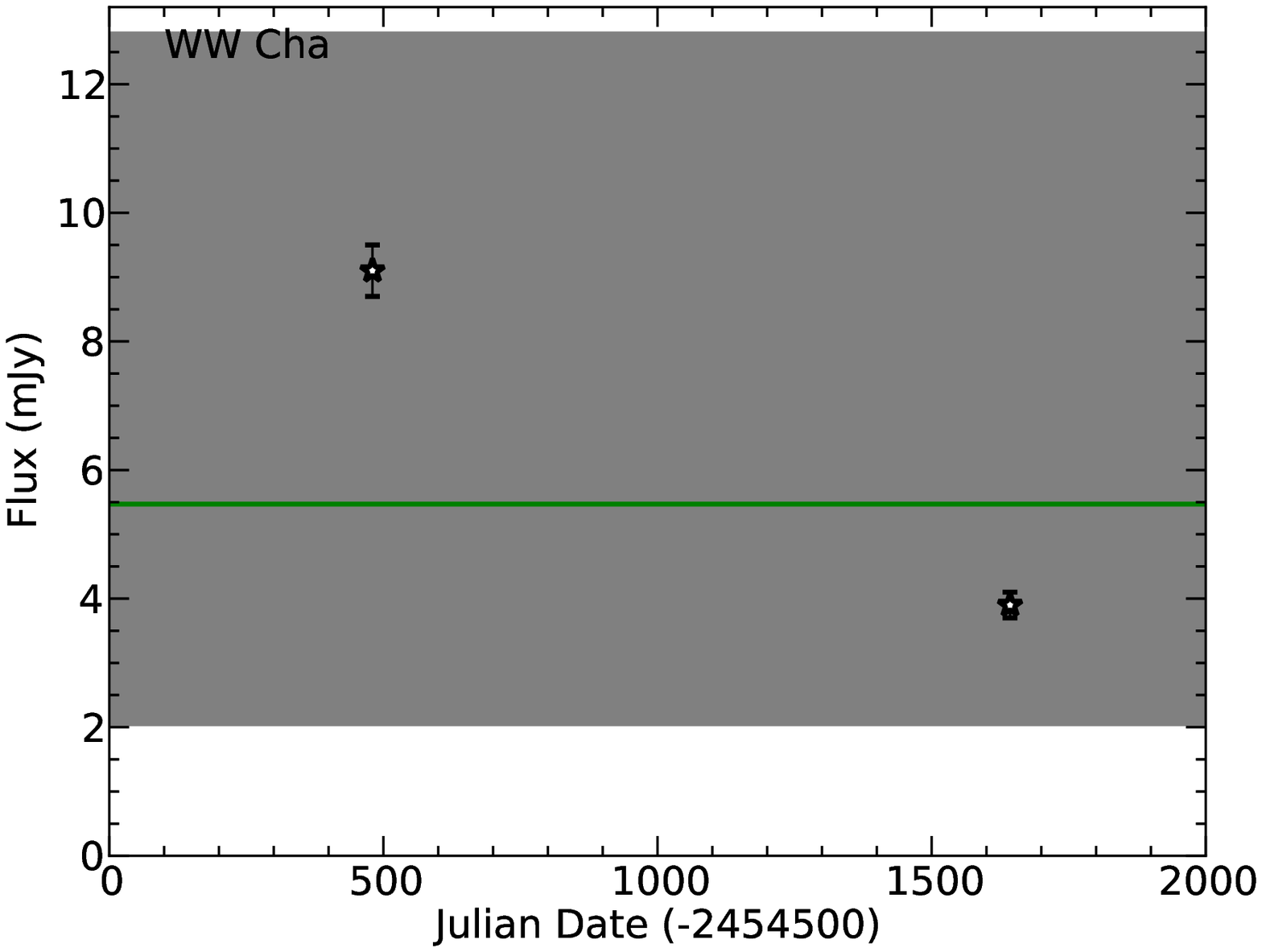}}
\caption[]{Continued.}
\label{fig-epochs}
\end{figure*}

\begin{figure*}	
\ContinuedFloat\centering 
      {\includegraphics[width= 0.80\columnwidth]{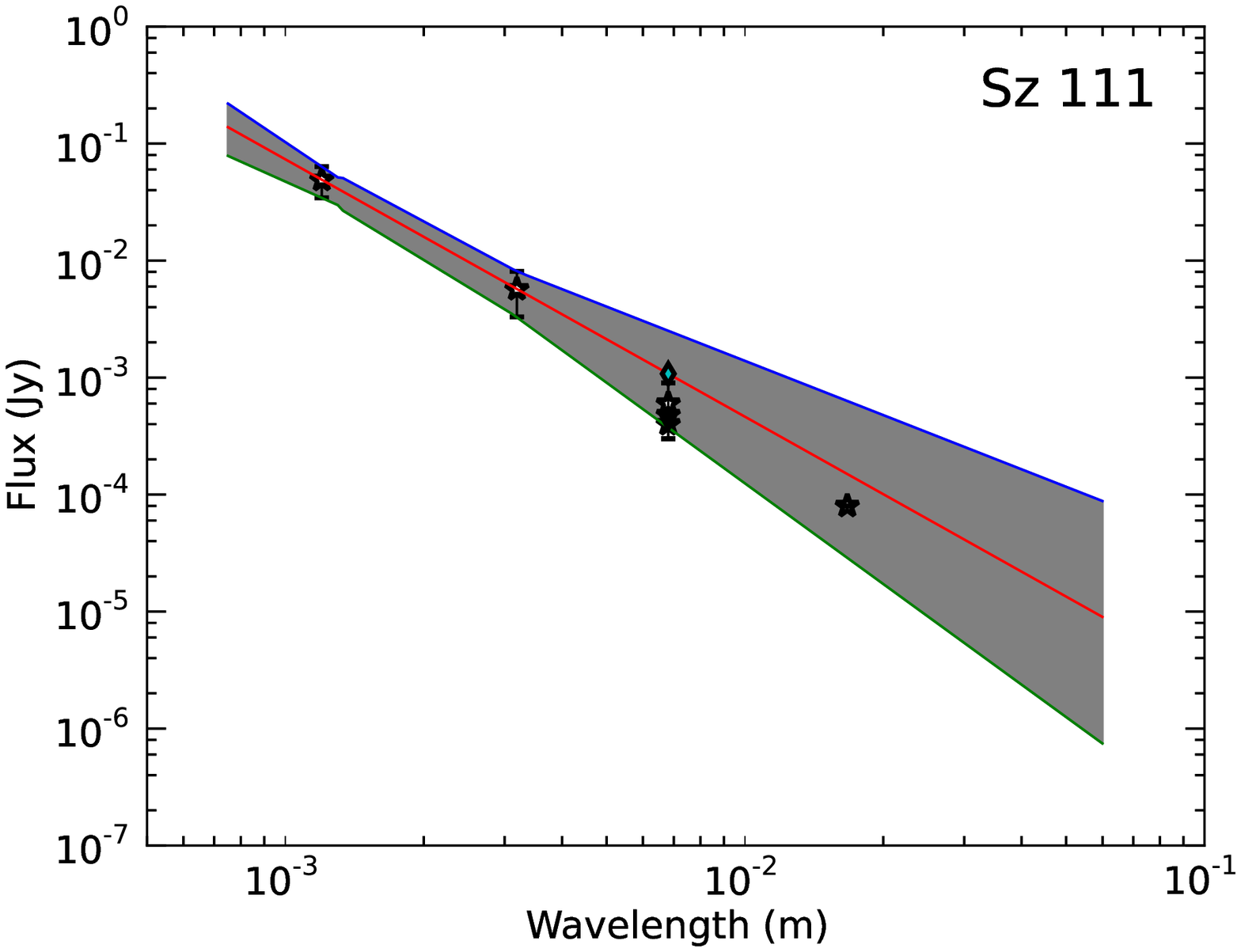}}\quad
      {\includegraphics[width=0.80\columnwidth]{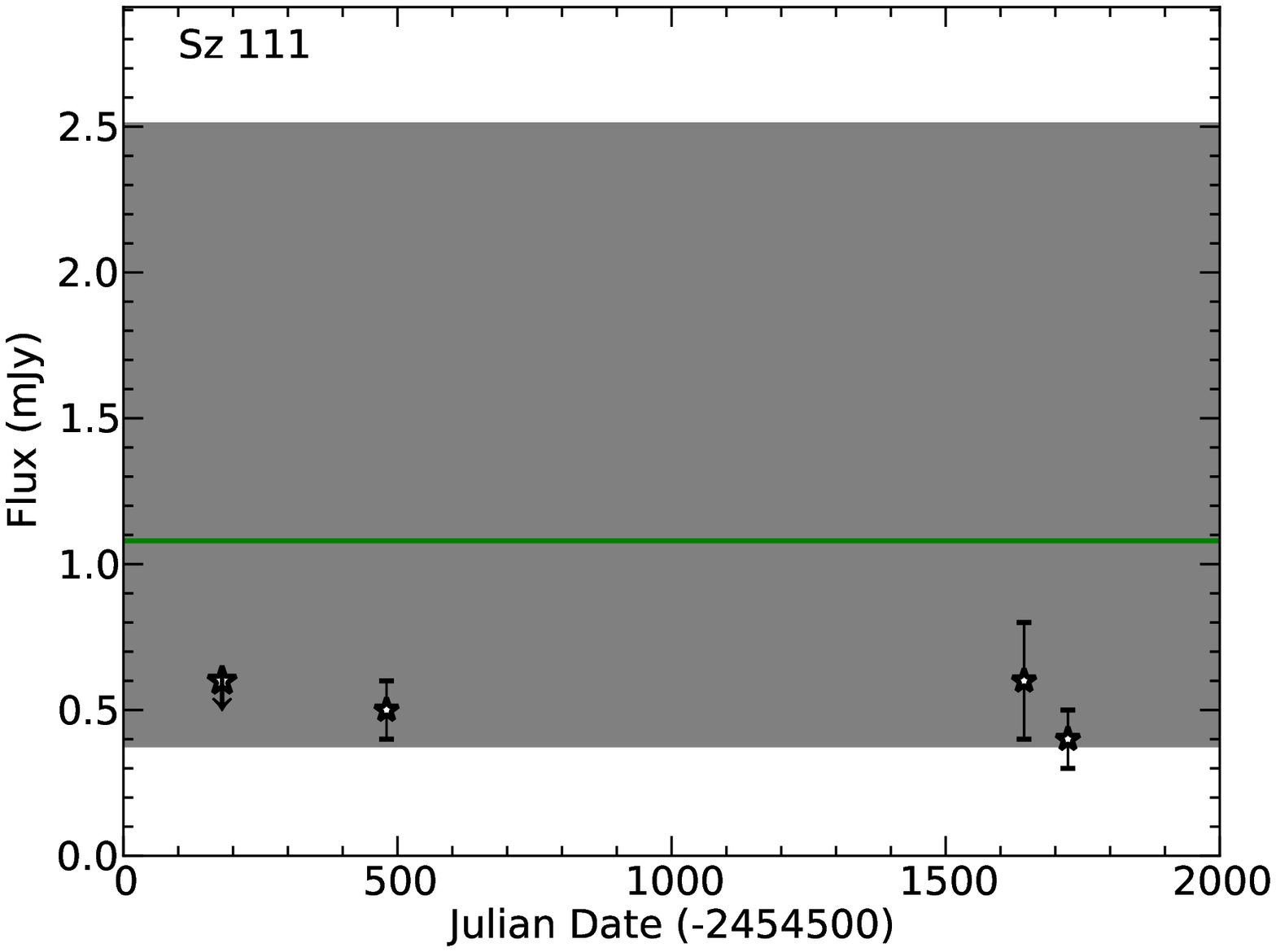}}\quad
      {\includegraphics[width= 0.80\columnwidth]{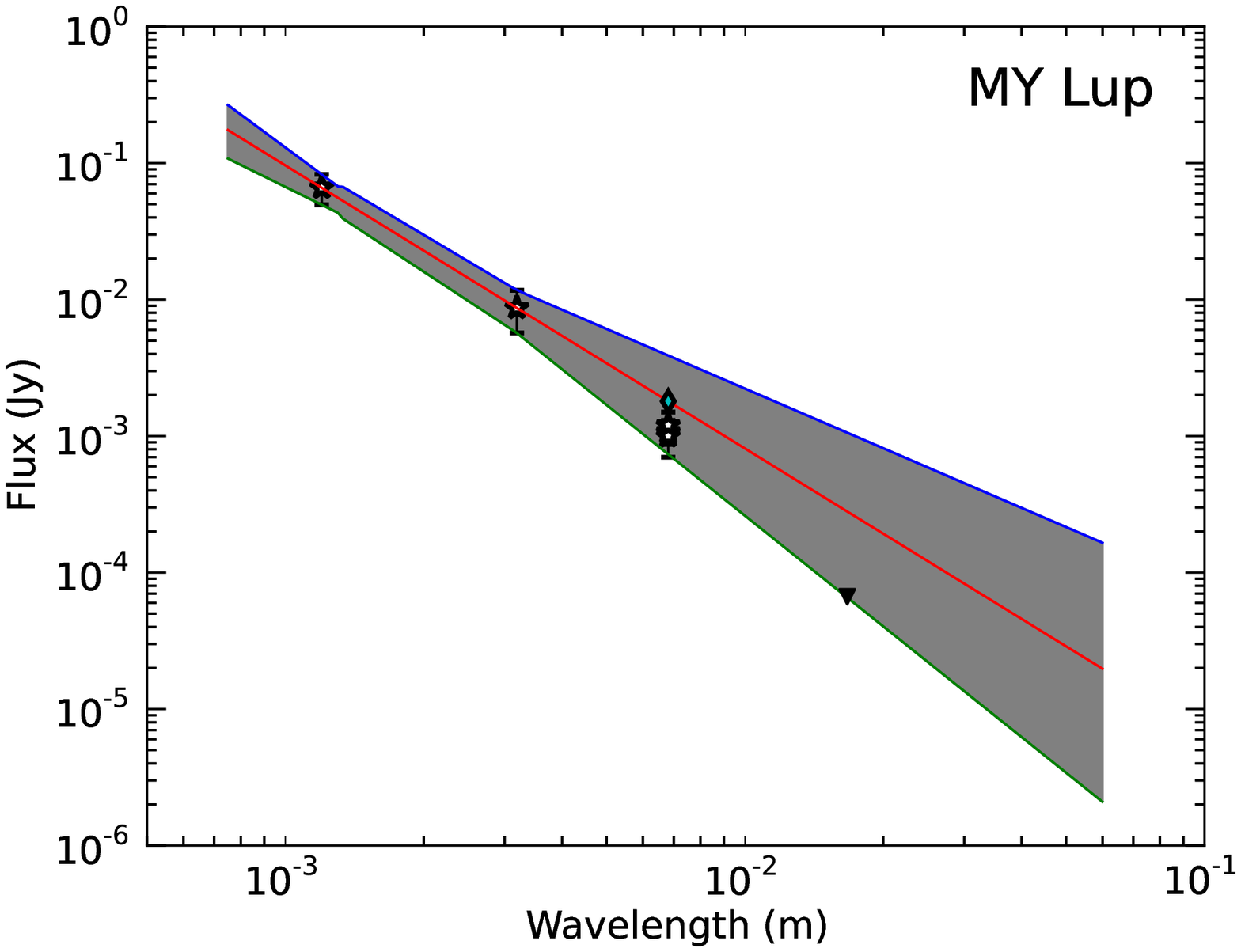}}\quad
      {\includegraphics[width=0.80\columnwidth]{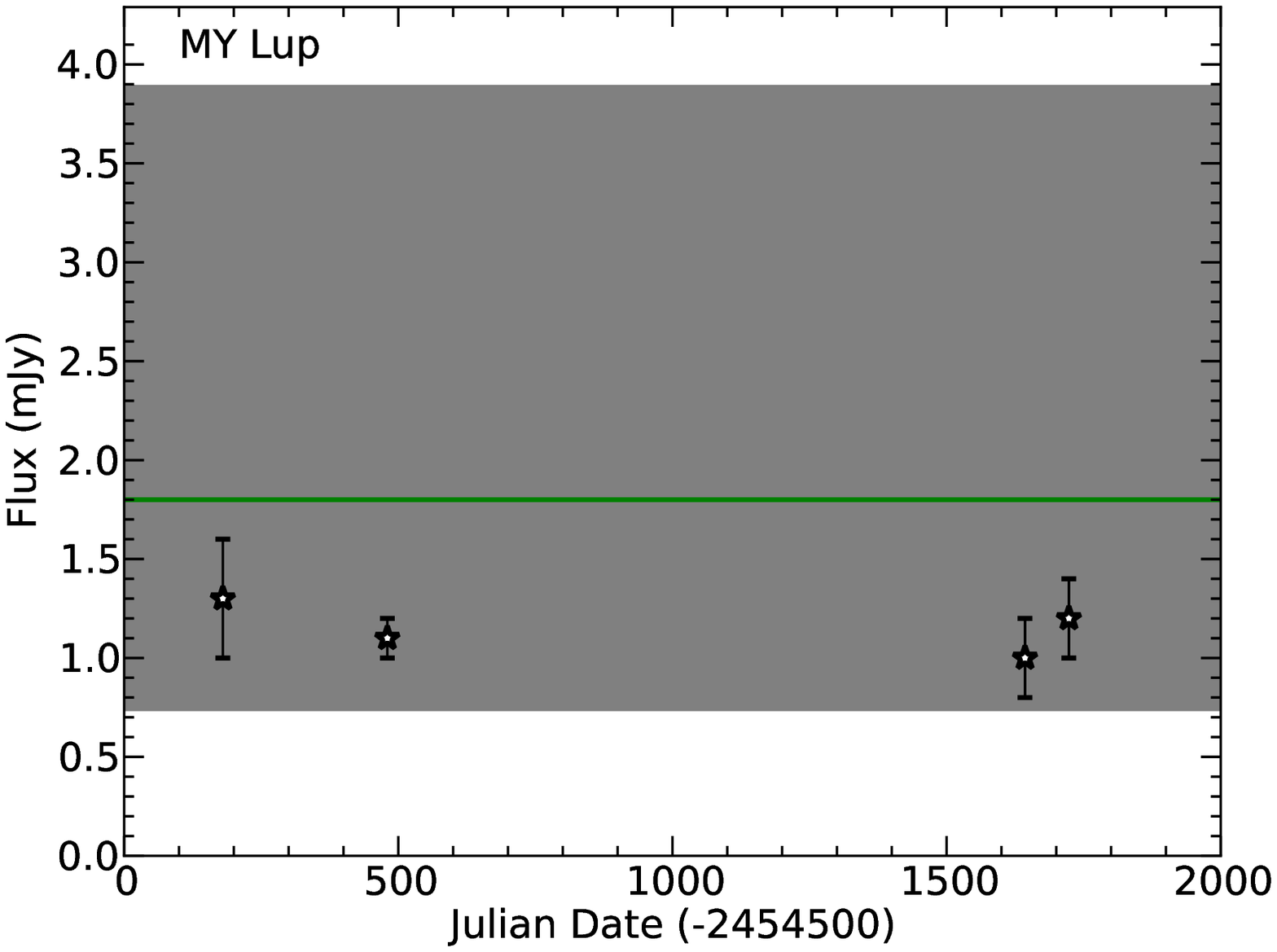}}\quad
      {\includegraphics[width= 0.80\columnwidth]{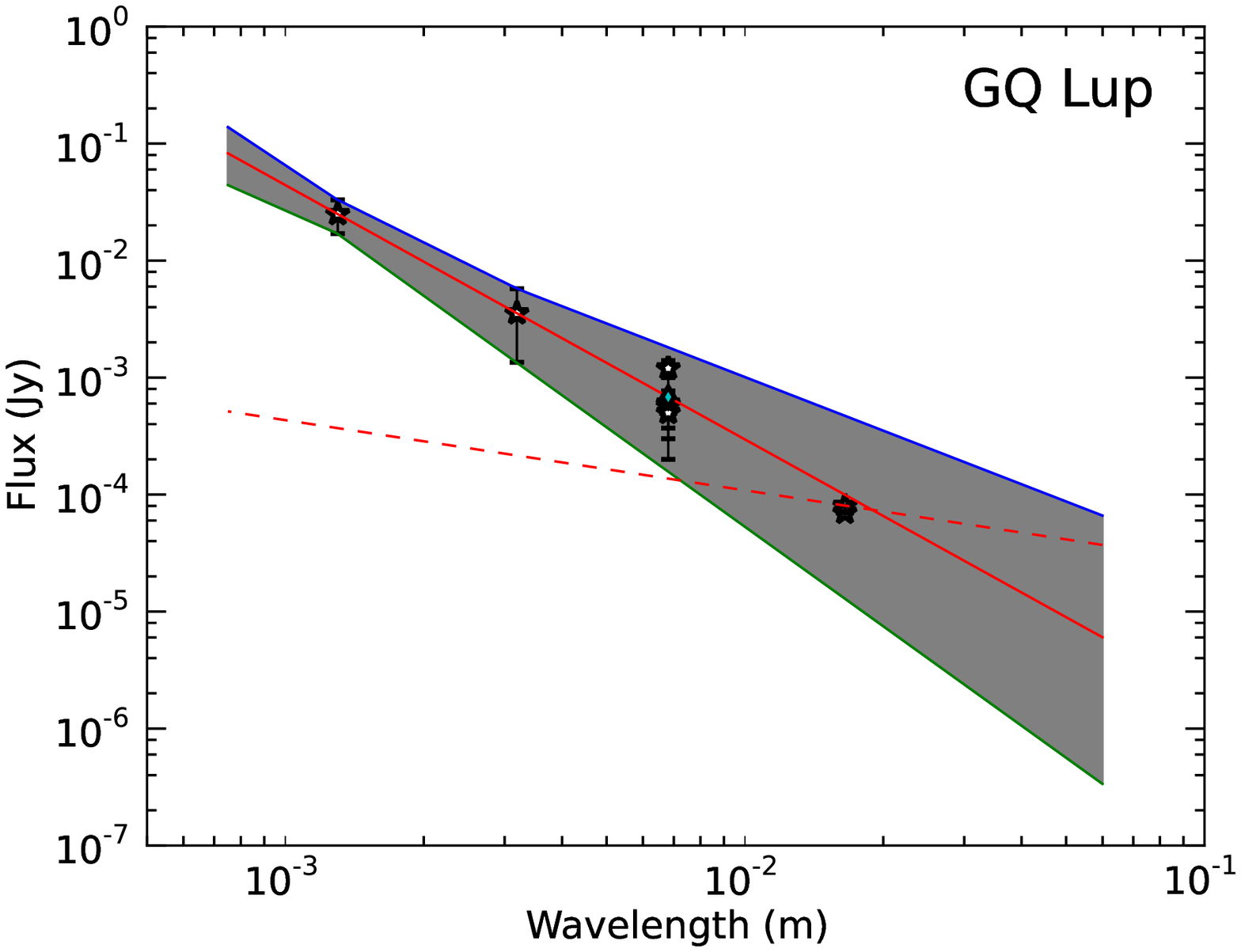}}\quad
      {\includegraphics[width=0.80\columnwidth]{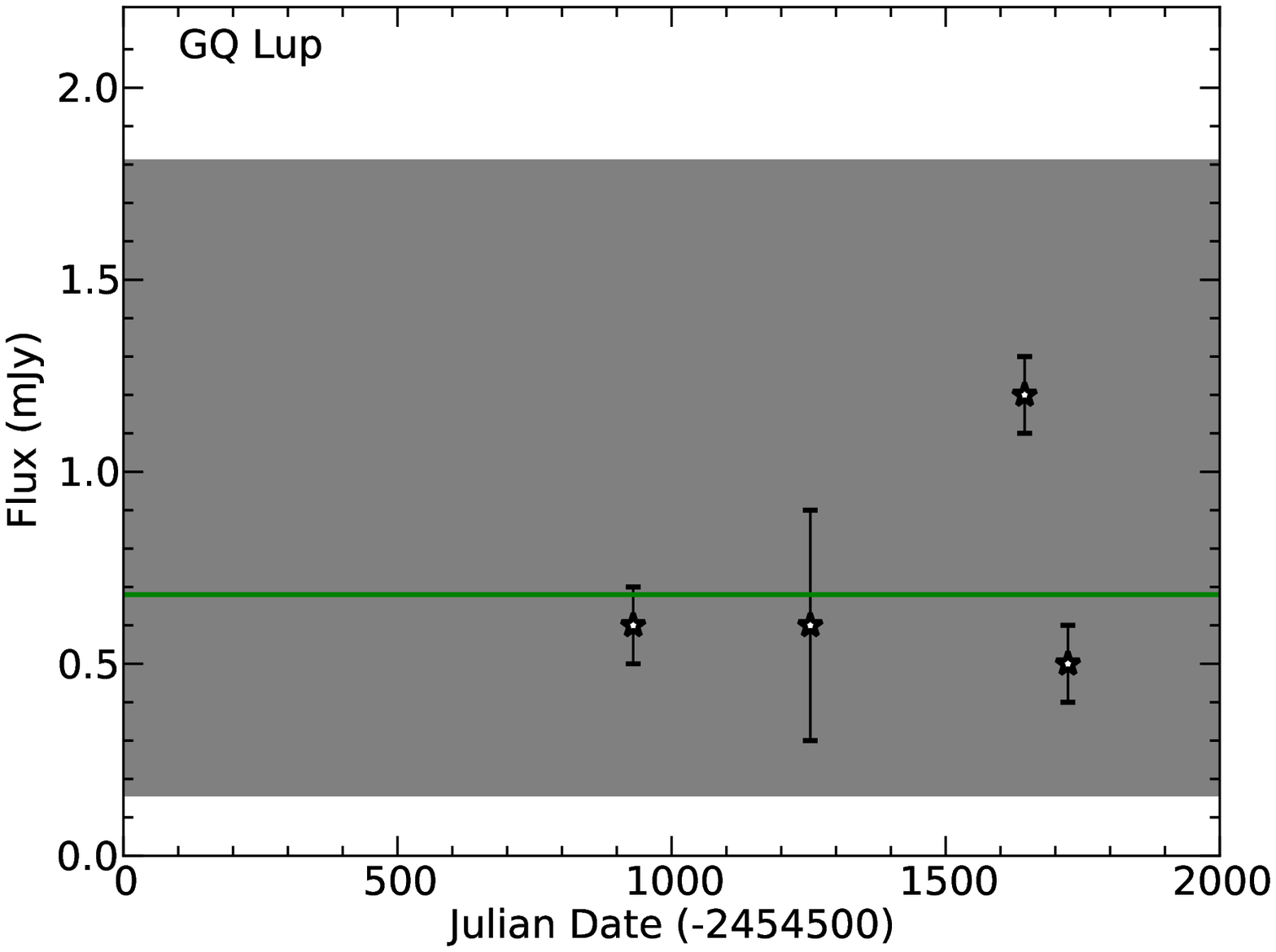}}
\caption[]{Continued.}
\label{fig-epochs}
\end{figure*}

 \begin{figure}
  	\centering 
   	\includegraphics[width=0.90\columnwidth]{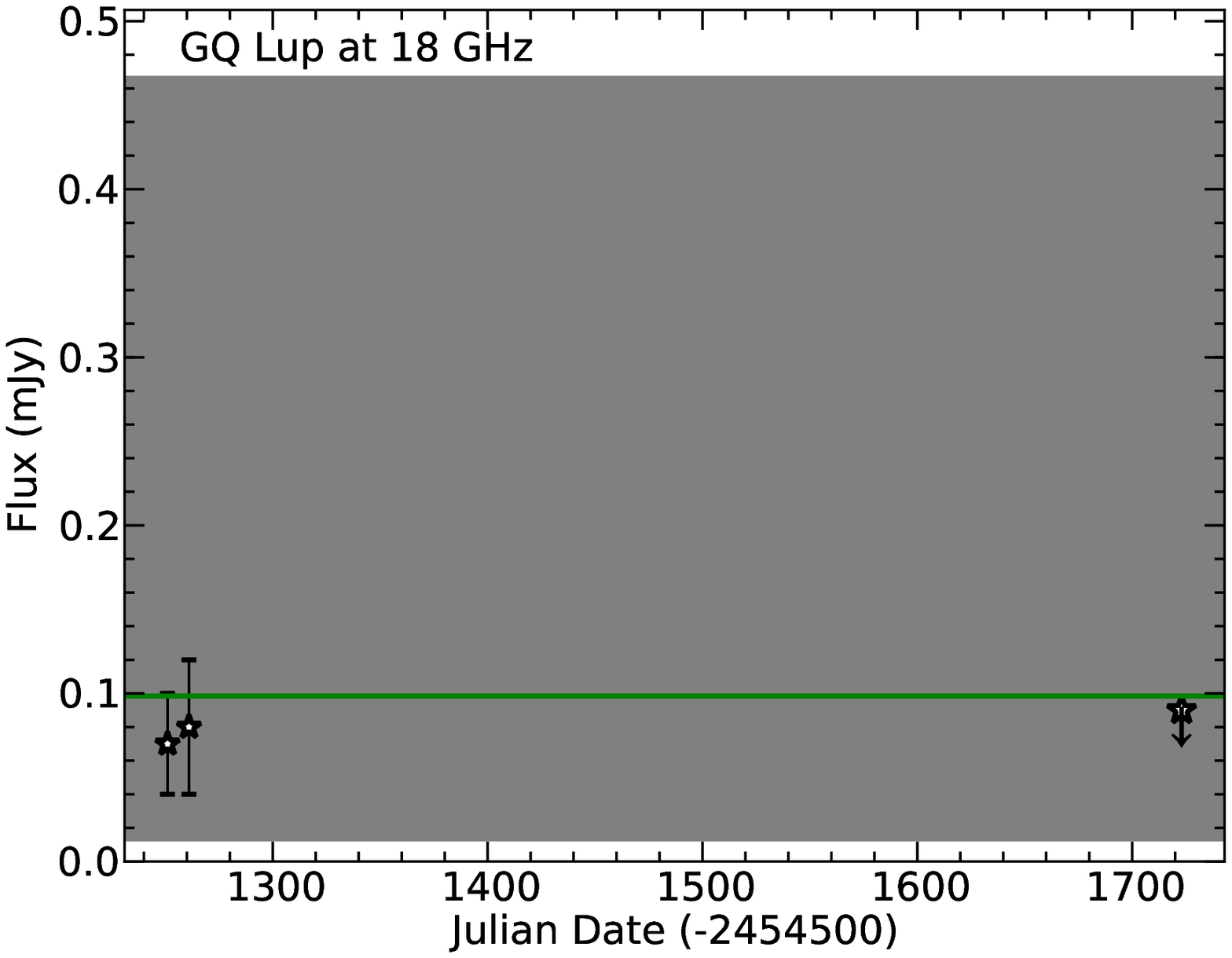}
    	\caption{The 15~mm continuum fluxes for GQ~Lup plotted from the first recorded epoch. The error bars represented the flux uncertainties. No flux variability was observed over a year timescale, suggesting the emission detected is primarily from thermal dust. The green diamond symbol/line correspond to the thermal dust emission component of the fluxes at 15~mm, error bars are the uncertainties for the estimated value. Continued in the next page }
    	\label{fig-epochs-gqlup}
 \end{figure}
 
\subsection[gg]{Grain growth}

{{For this sample of 11 protoplanetary discs, the two sources with no detected flux variability at 7~mm, MY~Lup and Sz~111, also have $\alpha<3$.  The four sources with high to extreme 7~mm fractional excess all have $\alpha >3$, while all sources with moderate to low fractional excess---with the exception of Glass~I--- have $\alpha<3$.}}
{{Applying the simplified \citet{draine06} relationship of $\beta\sim\alpha-2$, our results suggest that the majority of the sources have $\beta<1$, indicative of grain growth up to cm sizes, and the four most active sources with clear signs of excess emission above thermal dust have $\beta>1$, suggesting little grain growth. These results show that both variability and evidence for grain growth can be present for the same protoplanetary disc, but that grain growth appears inhibited in the most active sources, at least for this small sample.
}}

\subsection{Excess emission {at millimetre wavelengths}}
We have thus far determined that the presence of emission mechanisms other than thermal dust emission at 7~mm is common in protoplanetary discs, and that signatures of grain growth can be found in sources with and without 7~mm flux variability. 
{The 7~mm flux variability of $\sim30$ per cent over a period of 100s of days is consistent with the characteristics of thermal free-free emission.}
The presence of flux variability has also been detected at other wavelengths, such as in the X-ray, optical, and in the near-infrared (NIR). 

All the sources in our sample (with the exception of MY~Lup and Sz~111) have been detected in the X-ray with a $log(L_{x}/L_{bol})$ between $-3$ and $-4$, classifying these stars as X-ray active \citep[e.g.][]{2004A&A...423.1029S,2004ApJ...614..267F}.
A few sources in our sample have been studied more closely in the literature. 
Using the \textit{XMM-Newton} observations of CR~Cha, \citet{2006A&A...449..737R} detected a small flare and flux variability consistent with coronal activity and the presence of accretion. 
CS~Cha, T~Cha and GQ~Lup were included in \citet{2010A&A...519A.113G} analyses of \textit{Spitzer} [Ne II] line and \textit{XMM-Newton} archive data, where they classified CS~Cha as a jet-driving object with a transitional disc, T~Cha as a transitional disc and GQ~Lup as an optically thick disc without a known jet.
Similarly, X-ray flares have also been detected for Sz~32 and WW~Cha \citep{2004ApJ...614..267F}.
{{It is interesting to note {that }sources with flux variability at mm wavelengths are also variable in the X-ray; specifically the sources with high amplitude variability in the radio (CS~Cha, GQ~Lup, Sz~32, WW~Cha) are also very active in the X-ray. This suggests that the cause of the variability could be the same at both wavelengths. {Simultaneous observations at both wavelengths would be needed to confirm this relationship.}}}

Flux variability of up to 20 per cent is commonly found in young stellar objects in the optical and NIR wavebands \citep{2012MNRAS.420.1495S,2014MNRAS.443.1587R}. This variability can be caused by hot spots created by an accretion shock, variable circumstellar extinction, or a combination of these processes \citep{2012MNRAS.420.1495S}. Although variability above 20 per cent in the optical and NIR is rare, it has been documented. Additional optical images in the R and I bands and NIR images in the J and K bands have provided evidence that Sz~32 is variable at these wavelengths by more than the expected 20 per cent \citep{2014MNRAS.443.1587R}. The cause of the additional flux variability is still under investigation. 

\section{Conclusions}
\label{sec5-conclusions}

{{
We present a radio flux monitoring survey of 11 T~Tauri~stars in the Chamaeleon and Lupus Southern star forming regions over timescales of 100s of days. We found that the 7~mm flux varies by at least 30 per cent and up to a factor of 3 for most sources, indicating that processes other than just thermal dust are contributing to the emission. Flux monitoring of CS~Cha and GQ~Lup at 15~mm over the course of a year found consistent fluxes, suggesting no  variability over these timescales for these observations. 
Monitoring of T~Cha, Sz~32 and WW~Cha at 3 and 6~cm  also suggests excess emission above thermal dust. From our analysis of the radio spectrum and flux monitoring, only two sources show no variability and seem to be dominated by thermal dust emission: Sz~111 and MY~Lup.}}

{
Additionally, we looked at the flux variability in the X-ray, optical and near-infrared wavelengths, and found further support for the need for multiple epoch and multi-wavelength observations to determine the cause of the flux variability.}

We found seven sources with $\beta<1$, indicative of grain growth up to centimetre sized grains and four sources with a $\beta>1$, suggesting little to no grain growth. These results, with the exception of Sz~32, are consistent with reported values for spectral slope from 1--3~mm. Of the seven sources with signatures of grain growth, only CR~Cha and MY~Lup have no reported 7~mm flux variability. Thus, these results show that both signatures of grain growth and evidence of other emission mechanism at 7~mm may be present for the same protoplanetary disc. 

{
In the near future, the high resolution and sensitivity of Atacama Large Millimeter/Submillimeter Array (ALMA) will play an important role in answering these questions by significantly reducing the uncertainties in the 1~mm fluxes for the Chamaeleon sources, and thus reducing the uncertainties in the 1--3~mm spectral slope and $\beta$ values, and providing a better understanding of the signatures of grain growth in these protoplanetary discs. 
}

\section*{Acknowledgements}

We thank Hans Guenther and Scott Wolk for their help with the XMM-Newton Serendipitous Source Catalogue and for helpful discussions. 
{We also thank Leondardo Testi for useful discussions, and  the referee for their useful comments and constructive feedback.}
The Australia Telescope Compact Array is part of the Australia Telescope National Facility which is funded by the Commonwealth of Australia for operation as a National Facility managed by CSIRO. 
This research was supported in part by a Swinburne University Postgraduate Research Award and CSIRO OCE Postgraduate Top Up Scholarship. 
CMW acknowledges support from the Australian Research Council through Discovery Grant DP0345227 and Future Fellowship FT100100495.

\bibliographystyle{mnras}
\bibliography{abbrevs,biblio-2}

\appendix
\label{appendix1}

\section{Observing logs}
\label{ap-obs-log}

\begin{table*} 	
\caption{ATCA observing log for Lupus sources from this work and literature. (1) Source name. (2) Date of observations. (3) Frequency pair. (4) Flux calibrator. (5) ATCA array configuration. (6) Synthesized beam size. (7)  Total integration time used for analysis. (8) RMS. (9) Reference.}
 \footnotesize{
\begin{tabular}{llclccccl}
\hline \hline
	(1) & (2) & (3) & (4) &(5) & (6) & (7) & (8) &(9) \\
 	Source		&	Date		&	Freq. 	&	Flux	&		Antenna(s)	& Beam size &  \rm{T$_{\rm{int}}$}  & RMS &	Reference	\\
            	&	         	&	Pair (GHz)	&	Cal.	&	Config.	     &	 arcsec	&  (min)  & (mJy/Beam) &	\\ 	
   \hline\hline
Sz 111	&	Aug-08	&	43+45	&	Uranus	&	H214	&		$4$	&	---	&	0.2	&	\citet{Lommen10}	\\
	&	May-09	&	43+45	&	Uranus	&	H214	&		$5.2\times3.5$	&	60	&	0.1	&	\citet{Ubach1}	\\
	&	Aug-12	&	43+45	&	PKS~B1934-638	&	H75	&		$14.6\times10.3$	&	60	&	0.1	&	This work	\\
	&	Oct-12	&	43+45	&	PKS~B1934-638	&	H214	&		$5.0\times3.6$	&	60	&	0.03	&	This work \\
	&	Oct-12	&	17+19	&	PKS~B1934-638	&	H214	&	$13.1\times9.8$	 &	90	&	0.02	&	This work	\\
\hline
MY Lup	&	Aug-08	&	44.2+46.2	&	Uranus	&	H214	&		$4$	&		&	0.1	&	\citet{Lommen10}	\\
	&	May-09	&	43+45	&	Uranus	&	H214	&	$4.8\times3.4$	&	50	&	0.1	&	\citet{Ubach1}	\\
	&	Aug-12	&	43+45	&	PKS~B1934-638	&	H75	&		$14.8\times10.5$	&	50	&	0.1	& This work		\\
	&	Oct-12	&	43+45	&	PKS~B1934-638	&	H214	&		$4.7\times3.4$	&	80	&	0.03	&	This work	\\
	&	Oct-12	&	17+19	&	PKS~B1934-638	&	H214	&		$13.6\times9.7$	&	40	&	0.02	&	This work	\\
	\hline																											
GQ Lup	&	21-Aug-10	&	43+45	&	Uranus	&	H168	&	$6.7\times4.4$	&	80	&	0.04	&	\citet{Ubach1}	\\
	& 	10-Jul-11	&	43+45	&	Uranus	&	H214	&		$6.8\times3.6$	&	18	& 	0.07	&	\citet{Ubach1}	\\
	&	4-Aug-12	&	43+45	&	PKS~B1934-638	&	H75	&		$22.2\times9.2$	&	53	&	0.05	&	This work	\\
	&	22-Oct-12	&	43+45	&	PKS~B1934-638	&	H214	&		$5.0\times3.6$	&	30	&	0.04	&	This work	\\
	&	8-Jul-11	&	17+19	&	PKS~B1934-638	&	H214	&		$18.3\times9.7$	&	24	&	0.01	&	\citet{Ubach2014}	\\
	& 	18-Jul-11	&	17+19	&	PKS~B1934-638	&	H214	&		$17.8\times9.1$	&	20	& 	0.02	&	\citet{Ubach2014}	\\
	&	22-Oct-12	&	17+19	&	PKS~B1934-638	&	H214	&		$13.2\times9.4$	&	30	&	0.03	&	This work	\\
\hline
  \end{tabular}
  }	
   \label{tab-obs-log-lupus}	
  \end{table*}

 \begin{table*}
 \caption{ATCA observing log for Chamaeleon sources from this work and the literature. (1) Source name. (2) Date of observation. (3) Frequency pair. (4) Flux calibrator. (5) ATCA array configuration. (6) Synthesized beam size. (7)  Total integration time used for analysis. (8) RMS. (9) Reference.}
  \footnotesize{
 \begin{tabular}{llclccccl}
 \hline \hline
 	Source		&	Date		&	Freq. 	&	Flux	&		Antenna(s)	& Beam size&  \rm{T$_{\rm{int}}$}  & RMS &	Reference	\\
 	            &               &	Pair (GHz)	&	Cal.	&	Config.	&	 arcsec	  & (min)  &	(mJy/Beam) &\\
   \hline\hline
   CR~Cha	&	May-09	&	43+45	&	Uranus	&	H214	&			$5.0\times4.3$	&			80			&		0.1		&	\citet{Ubach1}	\\
   	&		Aug-12	&	43+45	&	PKS~B1934-638	&	H75	&			$16.3\times10.6$	&		20		&		0.1		&  This work	\\				
\hline
CS Cha	&	26-Apr-08	&	43+45	&	PKS~B1921-293	&	6A	&		$2.4\times1.4$	&	413	&	0.1	&	\citet{Lommen09}	\\
	&	05-Jul-08	&	43+45	&	Uranus	&	1.5B	&		$11.3\times0.6$	&	49	&	0.2	&	\citet{Lommen09}	\\
	&	06-Jul-08	&	43+45	&	Uranus	&	1.5B	&	$4.5\times0.6$	&	103	&	0.2	&	\citet{Lommen09}	\\
	&	May-09	&	43+45	&	Uranus	&	H214	&			$5.0\times4.3$ &			80			&	0.1		&	\citet{Ubach1}	\\
	&		Aug-12	&	43+45	&	PKS~B1934-638	&	H75	&		$15.9\times10.3$	&	20		&	0.1	& This work	\\
	&		Oct-12		&	43+45	&	PKS~B1934-638	&	H214	&			$5.4\times4.1$	&	80		&	0.3	&	This work	\\
	&		Jul-11	&	17+19	&	PKS~B1934-638	&	H214	&	$15.5\times12.6$	&	105		&	0.1	&	\citet{Ubach1}	\\
	&		Oct-12		&	17+19	&	PKS~B1934-638	&	H214	&			$13.9\times11.4$	&	126		&		0.03	&	This work	\\
\hline		
DI~Cha	&	May-09	&	43+45	&	Uranus	&	H214	&		$4.8\times4.2$	&	100	&	0.1	&	\citet{Ubach1}	\\
	&	Aug-12	&	43+45	&	PKS~B1934-638	&	H75	&		$13.8\times10.7$	&	30	&	0.1	&	This work	\\
\hline	
T~Cha	&	May-09	&	43+45	&	Uranus	&	H214	&		$5.1\times4.1$	&	60	&	0.1	&	\citet{Ubach1}	\\
	&	Aug-12	&	43+45	&	PKS~B1934-638	&	H75	&		$17.0\times10.4$	&	30	&	0.1	& This work	\\
	&	Jul-11	&	9	&	PKS~B1934-638	&	H214	&		$3.2\times1.1$	&	170	&	0.1	&	\citet{Ubach1}	\\
	&	Jul-12	&	9	&	PKS~B1934-638	&	H168	&		$5.3\times1.1$	&	20	&	0.05	&	This work	\\
	&	Jul-11	&	5.5	&	PKS~B1934-638	&	H214	&		$3.4\times1.9$	&	170	&	0.1	&	\citet{Ubach1}	\\
	&	Jul-12	&	5.5	&	PKS~B1934-638	&	H168	&		$8.9\times1.7$	&	20	&	0.08	& This work	\\
\hline		
Glass I	&	May-09	&	43+45	&	Uranus	&	H214	&	$4.9\times4.1$	&	150	&	0.1	&	\citet{Ubach1}	\\
	&	Aug-12	&	43+45	&	PKS~B1934-638	&	H75	&		$14.0\times11.4$	&	30	&	0.1	& This work	\\
\hline		
SZ Cha	&	May-09	&	43+45	&	Uranus	&	H214	&		$5.0\times4.1$	&	200	&	0.1	&	\citet{Ubach1}	\\
	&	Aug-12	&	43+45	&	PKS~B1934-638	&	H75	&	$14.1\times11.0$	&	40	&	0.1	&This work		\\
	&	Oct-12	&	43+45	&	PKS~B1934-638	&	H214	&		$5.4\times4.0$	&	100	&	0.1	&This work		\\
		&	Oct-12	&	17+19	&	PKS~B1934-638	&	H214	&		$13.7\times11.3$	&	160	&	0.02	& This work	\\
\hline	
Sz~32	&	31-March-08	&	43+45	&	QSO~B1057-797	&	H168	&		$5.8\times5.3$	&	119	&	0.2	&	\citet{Lommen10}	\\
	&	May-09	&	44+45	&	Uranus	&	H214	&		$5.0\times4.0$	&	130	&	0.2	&	\citet{Ubach1}	\\
	&	Aug-12	&	43+45	&	PKS~B1934-638	&	H75	&		$14.6\times11.3$	&	20	&	0.1	& This work		\\
	&	Jul-11	&	9	&	PKS~B1934-638	&	H214	&		$2.4\times1.8$	&	125	&	0.1	&	\citet{Ubach1}	\\
	&	Jul-12	&	9	&	PKS~B1934-638	&	H168	&	$1.5\times1.0$	&	20	&	0.05	&	This work	\\
	&	Jul-11	&	5.5	&	PKS~B1934-638	&	H214	&		$3.2\times2.0$	&	125	&	0.1	&	\citet{Ubach1}	\\
	&	Jul-12	&	5.5	&	PKS~B1934-638	&	H168	&	$3.3\times1.7$	&	20	&	0.04	&	This work	\\
\hline			
WW Cha	&	May-09$^\sigma$	&	43+45	&	Uranus	&	H215	&		$5.0\times4.0$	&	  	130		&	0.2	&	\citet{Ubach1}	\\
	&	Aug-12	&	43+45	&	PKS~B1934-638	&	H75	&		$17.4\times10.6$ 	&	20	&	0.1	& This work		\\
	&		Jul-11$^\sigma$	&	9	&	PKS~B1934-638	&	EW352	&		$2.4\times1.8$	&	125	&	0.1	&	\citet{Ubach2014}	\\
	&	Jul-12$^\sigma$	&	9	&	PKS~B1934-638	&	H168	&		$1.5\times1.0$	&	20	&	0.05	& This work		\\
	&		Jul-11$^\sigma$	&	5.5	&	PKS~B1934-638	&	EW352	&	$3.2\times2.0$	&	125	&	0.1	&	\citet{Ubach2014}	\\
	&	Jul-12$^\sigma$	&	5.5	&	PKS~B1934-638	&	H168	&	$3.3\times1.7$	&	20	&	0.04	& This work		\\							
 	\hline
  \end{tabular}
   }
   \label{tab-obs-log-cham}	
  \begin{tablenotes}
  	\footnotesize{ 
  		\item[2] $^\sigma$ WW Cha was not the primary target but was in the field of Sz~32.
  		}
  \end{tablenotes}	
  \end{table*}

\newpage

\section{Results at 15~mm}

 \begin{table*}
 \caption{Summary of single 15~mm band observations. (1) Source name. (2) Date of observation. (3) Total integration time used to determine flux. (4) Combined frequency. (5) Point flux (3$\sigma$ upper limit for non-detections). (6) RMS. (7) Beam size. (8) ATCA array configuration.}
 \centering
 \footnotesize{
 \begin{tabular}{llcccccl}
 \hline \hline
  (1) & (2) & (3) & (4) &(5) & (6) & (7) & (8) \\
  Source	&	Date	&	\rm{T$_{\rm{int}}$}	&	Freq. 	& 	Flux	& 	RMS	& 	Beam size & Array	\\
  	&		&	(minutes)	&	(GHz)	&	(mJy)	&	(mJy/beam)	&	(arcsecs) & Config.	\\
  \hline	\hline	
	SZ~Cha 	&	Oct-12	&	160	&	18	&	$0.14\pm0.03$	&	0.02	&	$13.7\times11.3$	&	H214	\\	
	Sz~111  	&	Oct-12	&	90	&	18	&	$0.08\pm0.02$	&	0.02	&	$13.1\times9.8$	&	H214	\\
	MY~Lup 	&	Oct-12	&	40	&	18	&	$<0.06$	&	0.02	&	$13.6\times9.7$	&	H214	\\
 \hline
 \end{tabular}
  }
  \label{tab-results-extras}	
 \end{table*}

\newpage

\section{Summary of WW~Cha Radio Monitoring}
\label{wwcha-obs-log}

\begin{table*}
 \caption{Summary of results for WW~Cha monitoring. (1) Source name. (2) Observation date. (3) Total integration time used to determine flux. (4) Frequency. (5) Point flux ($3\sigma$ for non-detections). (6) RMS. (7) Beam size. (8) ATCA array configuration. (9) References.}
 \centering
 \footnotesize{
 \begin{tabular}{llcccccll}
 \hline \hline
 (1) & (2) & (3) & (4) &(5) & (6) & (7) & (8) & (9) \\
  Source	&	Date	&	\rm{T$_{\rm{int}}$}	&	Freq. 	& 	Flux	& 	RMS	& 	Beam size & Array & References	\\
  	&		&	(minutes)	&	(GHz)	&	(mJy)	&	(mJy/beam)	&	(arcsecs) & Config. &	\\
  \hline	\hline
 WW Cha	&	5 Oct 2007	&		81.0	&	40.8	&	3.93$\pm$0.29	&	0.35	&	$11\times11$	&	 H75C	 & \citet{Lommen07} \\
	&	31 Mar 2008	&		119.4	&	40.8	&	5.19$\pm$0.17	&	0.194	&	$6.1\times5.6$	&	H168 & \citet{Lommen07}	\\
 	&	5 Oct 2007	&		81.0	&	42.5	&	5.41$\pm$0.32	&	0.313	&	$11\times11$	&	H75C & \citet{Lommen07}	\\
 	&	31 Mar 2008	&		119.4	&	42.5	&	5.10$\pm$0.17	&	0.231	&	$5.8\times5.3$	&	H168 & \citet{Lommen07}	\\
	&	May-09	&		130		&	44	&	$3.4\pm0.1$	&	0.2	&	$5.0\times4.0$	&	H215 & \citet{Ubach1}	\\
 	&	Aug-12	&		20.0	&	44	&	$3.9\pm0.2$	&	0.1	&	$17.4\times10.6$	&	H75 & This work	\\
	&	Jul-11	&		70.0	&	18	&	$0.75\pm0.02$	&	0.1	&	$15.6\times12.3$	&	H214 & \citet{Ubach1}	\\
 	&	8 May 2006	&		48.0	&	18.4	&	$1.04\pm0.30$	&	0.158	&	$23\times6$	&	H214C & \citet{Lommen07}	\\
 	&	13 Oct 2006	&		78.6	&	18.4	&	1.08$\pm$0.21	&	0.217	&	$26\times9$	&	H214C & \citet{Lommen07}	\\
 	&	18 Oct 2006	&		199.2	&	18.4	&	1.23$\pm$0.19	&	0.11	&	39$\times$7	&	EW352 & \citet{Lommen07}	\\
 	&	24 Oct 2007	&		189.0	&	18.4	&	0.95$\pm$0.13	&	0.181	&	13 $\times$ 9	&	H214C & \citet{Lommen07}	\\
 	&	2 Nov 2007	&		508.8	&	18.4	&	0.48$\pm0.13^c$	&	0.089	&	2.0 $\times$ 1.4	&	1.5A & \citet{Lommen07}	\\
 	&	31 Mar 2008	&		79.8	&	18.4	&	1.07$\pm$0.10	&	0.227	&	15 $\times$ 13	&	H168 & \citet{Lommen07}	\\
 	&	8 May 2006	&		48.0	&	18.5	&	1.16$\pm$0.31	&	0.329	&	23 $\times$ 6	&	H214C & \citet{Lommen07}	\\
 	&	13 Oct 2006	&		78.6	&	18.5	&	$<0.921$	&	0.307	&	$25\times9$	&	H214C & \citet{Lommen07}	\\
 	&	18 Oct 2006	&		199.2	&	18.5	&	0.81$\pm$0.20	&	0.222	&	38 $\times$ 7	&	EW352 & \citet{Lommen07}	\\
 	&	24 Oct 2007	&		189.0	&	19.4	&	1.01$\pm$0.16	&	0.209	&	12 $\times$ 8	&	H214C & \citet{Lommen07}	\\
 	&	2 Nov 2007	&		508.8	&	19.4	&	0.55$\pm0.20^c$	&	0.125	&	1.9 $\times$ 1.3	&	1.5A & \citet{Lommen07}	\\
 	&	31 Mar 2008	&		79.8	&	19.4	&	0.88$\pm$0.12	&	0.232	&	14 $\times$ 13	&	H168 & \citet{Lommen07}	\\
	&	Jul-11	&		125.0	&	9.9	&	$<0.3$	&	0.1	&	$2.4\times1.8$	&	EW352 & \citet{Ubach1}	\\
	&	Jul-12	&		20.0	&	9.9	&	$<0.15$	&	0.05	&	$1.5\times1.0$	&	H168 & This work	\\
 	&	18 Oct 2006	&		237.0	&	8.6	&	$<0.222$	&	0.074	&	$71\times20$	&	EW352 & \citet{Lommen07}	\\
 	&	9 June 2007	&		311.4	&	8.6	&	0.63$\pm$0.06	&	0.076	&	$53\times16$	&	EW352 & \citet{Lommen07}	\\
	&	Jul-11	&		125.0	&	5.5	&	$<0.24$	&	0.1	&	$3.2\times2.0$	&	EW352 & \citet{Ubach1}	\\
	&	Jul-12	&		20.0	&	5.5	&	$<0.12$	&	0.04	&	$3.3\times1.7$	&	H168 & This work	\\
 	&	18 Oct 2006	&		237.0	&	4.8	&	$<0.202$	&	0.067	&	$125\times32$	&	EW352 & \citet{Lommen07}	\\
 	&	9 June 2007	&		311.4	&	4.8	&	$<0.399$	&	0.133	&	$91\times26$	&	EW352 & \citet{Lommen07}	\\
 	\hline
  \end{tabular}
   }
\label{tab-results-wwcha}	
\end{table*}

\label{lastpage}
\end{document}